\documentclass[aps,prb,twocolumn,showpacs,superscriptaddress,bibliography,notitlepage]{revtex4-1}
\usepackage[colorlinks=true,citecolor=blue,urlcolor=blue,linkcolor=blue,breaklinks=true]{hyperref}
\usepackage{amssymb}
\usepackage{graphicx}
\usepackage{amsmath}
\usepackage{mathrsfs}
\usepackage{mathtools}
\usepackage[export]{adjustbox}
\usepackage{epsfig}
\usepackage{times}
\usepackage{color}
\usepackage{setspace}
\usepackage{calc}
\usepackage{natbib}
\usepackage{bm}
\DeclareMathAlphabet{\mathpzc}{OT1}{pzc}{m}{it}
\usepackage{array}
\usepackage{multirow}
\usepackage{comment}
\usepackage{float}
\usepackage{soul}

\begin{document}

\newcommand {\beq} {\begin{equation}}
\newcommand {\eeq} {\end{equation}}
\newcommand {\bqa} {\begin{eqnarray}}
\newcommand {\eqa} {\end{eqnarray}}
\newcommand {\bseq} {\begin{subequations}}
\newcommand {\eseq} {\end{subequations}}
\newcommand {\baln} {\begin{align}}
\newcommand {\ealn} {\end{align}}
\newcommand{\dhat}{\ensuremath{\hat{D}}}
\newcommand{\ehat}{\ensuremath{\hat{E}}}
\newcommand{\lhat}{\ensuremath{\hat{\Lambda}}}
\newcommand{\zbar}{\ensuremath{\bar{\zeta}}}
\newcommand{\ebar}{\ensuremath{\bar{\eta}}}
\newcommand {\ba} {\ensuremath{b^\dagger}}
\newcommand {\Ma} {\ensuremath{M^\dagger}}
\newcommand {\psia} {\ensuremath{\psi^\dagger}}
\newcommand {\psita} {\ensuremath{\tilde{\psi}^\dagger}}
\newcommand{\lp} {\ensuremath{{\lambda '}}}
\newcommand{\A} {\ensuremath{{\bf A}}}
\newcommand{\Q} {\ensuremath{{\bf Q}}}
\newcommand{\kk} {\ensuremath{{\bf k}}}
\newcommand{\qq} {\ensuremath{{\bf q}}}
\newcommand{\kp} {\ensuremath{{\bf k'}}}
\newcommand{\rr} {\ensuremath{{\bf r}}}
\newcommand{\rp} {\ensuremath{{\bf r'}}}
\newcommand {\ep} {\ensuremath{\epsilon}}
\newcommand{\nbr} {\ensuremath{\langle ij \rangle}}
\newcommand {\no} {\nonumber}
\newcommand{\up} {\ensuremath{\uparrow}}
\newcommand{\dn} {\ensuremath{\downarrow}}
\newcommand{\rcol} {\textcolor{red}}
\newcommand{\bcol} {\textcolor{blue}}
\newcommand{\bu} {\bold{u}}
\newcommand{\tr}[1]{\mathrm{Tr}\left[#1\right]}
\newcommand{\ve}[1]{\boldsymbol{#1}}
\newcommand{\args}[1]{\ve{#1},\ve{\bar{#1}}}
\newcommand{\mes}[1]{\mathcal{D}\hspace{-2pt}\left[\args{#1}\right]}
\newcommand{\ii}{\iota}
\newcommand{\mnm} {\ensuremath{\mathbb{M}}}
\newcommand{\ncr}[2]{\begin{pmatrix}#1\\#2\end{pmatrix}}
\begin{abstract}
We study the quench and the ramp dynamics of interacting $N$-component charged bosons
with dipole symmetry using Schwinger-Keldysh field theory in the large-$N$ limit. The equilibrium phase
diagram of these bosons shows two phases in the
large-$N$ limit. The first is a normal phase where both the global
 $U(N)$ and the dipole symmetries are conserved and the second is a
delocalized condensed phase where both the symmetries are broken. In
contrast, our explicit computation of the steady state after an instantaneous quantum quench from the condensed phase shows that an additional, novel, delocalized normal
phase, where the global $U(N)$ symmetry is conserved but the dipole
symmetry is broken, can exist for a range of quench parameters. A study of ramp dynamics of the model shows that
the above-mentioned steady state exists only above a critical ramp
rate which we estimate.

\end{abstract}

\title{Non-equilibrium dynamics of bosons with dipole symmetry: Large-$N$
Keldysh approach}
\author{Md Mursalin Islam}\email{mursalin@theory.tifr.res.in}
 \affiliation{Department of Theoretical Physics, Tata Institute of Fundamental
 Research, Mumbai 400005, India.}
\author{K. Sengupta}
\affiliation{School of Physical Sciences, Indian Association for the
Cultivation of Science, Kolkata 700032, India.}
\author{Rajdeep Sensarma}
\affiliation{Department of Theoretical Physics, Tata Institute of Fundamental
 Research, Mumbai 400005, India.}

\date{\today}

\maketitle
\section{Introduction}

The behaviour of systems with symmetries
that lead to conservation of multipole moments of charges (e.g.
dipole moment) has recently become a question of paramount interest
in the study of quantum many-body
systems~\cite{fracton_gauge,largeN_fracton,dipolehydrodynamics,fractonsf2,gorantla_seiberg1,kim1,nandkishore_kim,senthil1,PhysRevB.105.205127,PhysRevB.107.195131,PhysRevB.107.195132}.
 These symmetries which implement, for
example, dipole conservation are intermediate between global and
gauge symmetries. It is well-known that field theories with
this kind of symmetries often show a strong ``UV-IR'' mixing (i.e. low
energy properties like ground state degeneracy are sensitive to high
energy regularization of the theory)~\cite{gorantla_seiberg2,seiberg1,seiberg2,xichen1,slagle1,slagle_kim}. These systems are also the starting point for
describing matter coupled to tensorial gauge
fields~\cite{fracton_gauge,PretkoU1,gorantla_seiberg1}, which has
found applications in understanding fractonic phases of matter.
Such fractonic phases are characterized by low-energy excitations which are localized and can move only by creating additional excitations. In some cases, the excitations are localized in certain directions, while they are free to move in other directions. Their motion can thus be restricted along a line or a plane in three dimensional (3D) system~\cite{fracton_Haah,fracton_VijayHaahFu,PretkoU1,seiberg3,fractons,fractonphases}.

It is generally believed that the phase diagram of interacting
quantum matter out of thermal equilibrium can show a wider variety
of phases than systems are in thermal equilibrium. These
``phases'' can be true steady states in driven dissipative systems,
where the energy balance between the drive and the dissipation
determines the long time behaviour~\cite{drive_th1,drive_th2,drive_th3,drive_rev1,drive_rev2,drive_exp1}.
In fact there are protocols that engineer the bath and the
drive~\cite{zoller_bath1,zoller_bath2,zoller_bath3,bath_mbl,cirac_bath1} to guide quantum systems into non-thermal
states. An extreme example is the formation of the so-called
$\eta$-paired state of the Hubbard model in non-equilibrium
dynamics~\cite{etapair_yang,etapair_bernevig}. This state is a maximum of
the free energy and does not occur in thermal phases of the Hubbard
model~\cite{etapair_demler1}. Moreover, steady states in
closed quantum systems which violate eigenstate thermalization
hypothesis (ETH)\cite{eth1_Deutsch,eth2_Srednicki,eth3_Srednicki,eth4_Rigol,eth5_Reimann} are known to be athermal; such states occur in
integrable systems~\cite{integ_rigol,gge_gautam,integ_exp} or many-body
localized systems~\cite{mbl_th1,mbl_imbalance_expt,mbl_exp2,mbl_rev1,mbl_rev2,mbl_rev3,cgs_memory} and
carry the imprint of the initial conditions. They can show widely
different behaviour from thermal states. Similarly, in periodically driven
systems the ``phase'' represented by such steady states occur in a
prethermal regime
~\cite{prethermalization,pretherm_exp1,pretherm_exp2,pretherm_th1}.
Such prethermal states can persist for a long, experimentally
relevant timescale, and may show several interesting features
\cite{tc1,tc2,aba1,ks1,ks2,ks3, Marino_2022}. These prethermal states can also show
very different characteristics than the corresponding thermal states.

An important question that can be asked is the following: How can
one produce a phase (either as a steady state or as a pre-thermal
state), which has different symmetry content than the thermal phases
of the system? Is this possible through a simple quench or a ramp of
a parameter of the Hamiltonian, or does one necessarily need a more
fine-tuned protocol for it? In this work, we show (using a Keldysh formalism\cite{kamenevbook, realscalar} and in the large-$N$
limit) that interacting charged bosons with a dipole symmetry (that
conserves dipole moment), represented by a
$N$-component scalar field, can produce a steady state after an
instantaneous quantum quench which has different symmetries from its
equilibrium phases. In equilibrium, the system is either localized
or Bose condensed, while the non-equilibrium steady state can
support a non-condensed but dispersive phase with broken dipole
symmetry for certain range of quench parameters. If the parameters
are changed with a rate, this new dynamic phase disappears below a critical ramp-rate, thus reproducing the equilibrium
phase diagram as one approaches the adiabatic limit.

To study this phenomenon, we consider a field
theoretic model of $N$ flavours of complex bosons in three spatial
dimensions with dipole symmetry first proposed by Pretko
~\cite{fracton_gauge}. Here the bosons do not have a microscopic
kinetic energy and have strongly momentum-dependent interactions;
these features are consistent with global $U(N)$ charge and the
corresponding dipole moment conservation. Regularizing the theory on
a hyper-cubic lattice, we study its equilibrium and non-equilibrium
phase diagram in the large-$N$ limit. In equilibrium, the standard
$U(N)$ model, which does not have dipole conservation, has a continuous
quantum phase transition in 3D as a function of
the mass and the repulsive coupling in the theory. A key question
then comes up: if one enhances the symmetries to include dipole
conservation, what happens to this phase diagram? The dipole conservation implies an absence of bare kinetic energy term in the model; this leads to a very
different equilibrium phase diagram at zero temperature in the large-$N$ limit. At large values of mass, the system is in a normal phase
where both the $U(N)$ charge and the dipole moment is conserved. The
particles in this phase are completely localized in space. As the
mass is reduced the system undergoes a phase transition to a
condensed phase where the $U(N)$ symmetry gets broken down to $U(N-1)$. The dipole
symmetry is also broken in this phase leading to gapless  delocalized
excitations with finite dispersion. These two
phases are separated by a direct first-order transition. The
possible phase where $U(N)$ is conserved, but dipole symmetry is
broken, does not show up in the equilibrium phase diagram. We have
also considered the phase diagram at a finite temperature and it is
qualitatively similar to the ground state. Note that this phase diagram is qualitatively different from the phase diagram obtained in Ref.~\onlinecite{largeN_fracton} for Schr\"{o}dinger bosons with dipole conservation.

Interestingly, we find that the dynamical phase diagram of this
system, studied within a large-$N$ Keldysh formalism
\cite{kamenevbook,kamenevarticle,realscalar,cgs}, has a qualitatively different structure than its
equilibrium counterpart. We follow the fate of the system starting
from the ground state deep in the condensed phase, after an
instantaneous quantum quench of the bare mass using a formalism
developed in Ref.~\onlinecite{realscalar}. At long times, the
system goes into a quasi-stationary state, with observables
oscillating around an average value. One can get an effective static
description if one coarse grains the description on the scale of the
time period of these oscillations. As a function of the final bare
mass after the quench, we find three distinct steady states: (i) If
the quench is deep into the condensed phase, the steady state has a
finite condensate and breaks both $U(N)$ and dipole symmetry. (ii)
As the final bare mass is increased, the condensate in the steady
state continuously goes to zero at a critical value of the quenched
mass. Beyond this point the system shows a gapped spectrum, with the
gap increasing from zero continuously as a function of the quenched
mass.  In this phase, the dipole symmetry remains
broken and the spectrum is dispersive, i.e. we get a steady state
where the dipole symmetry is broken but $U(N)$ symmetry is restored.
This phase, which has no equilibrium analogue, shows
up as a steady state following a quench and is separated from the
condensed phase by a second order dynamic transition. (iii) As the
quenched mass is increased further, the gap shows a sharp jump at a
second critical value of the quenched mass, where the curvature of
the dispersion jumps from a finite value to zero. Thus, the system
undergoes a first-order transition to a state where dipole symmetry
is restored and the excitations are all localized.
This indicates that the single first-order transition of the
equilibrium state splits into a second order transition related to
breaking of $U(N)$ symmetry and a separate first-order transition
related to breaking of dipole symmetry.

We characterize the oscillations in the quasi-stationary state at long times. We find that the oscillation frequency scales with the gap  while the amplitude increases with the decreasing value of the curvature of the dispersion. We explain this in terms of the dynamical equations. Given the system always has oscillation in the dynamics, we look at the lowest values of the parameters to determine the transition.

We show that the presence of the phase with conserved $U(N)$ but
broken dipole symmetry depends on the initial state from which the
quench taken place. It vanishes when the initial
state is not deep in the condensed phase, i.e. beyond a critical value of the initial mass.  Beyond
this value of the initial mass, the system undergoes a first-order
transition similar to that found in equilibrium. This indicates that
the presence of the intermediate phase depends on energy pumped into
the system by the quench. This energy, pumped into the system, can
be tuned by using a ramp protocol which changes the mass at a
controlled rate allowing one to interpolate between the quench and
the adiabatic limits. We find that the dipole broken $U(N)$
conserved phase vanishes below a critical ramp rate. For ramps
slower than this critical rate, the system once again shows a single
first-order transition where both $U(N)$ and dipole symmetry is
broken simultaneously. We note that while the presence of the new dynamic phase
depends on the energy dumped into the system, one cannot simply
explain this phase in terms of thermalization at a finite energy
density, since this phase does not occur in the thermal phase diagram
of the system. The novel steady state is a new athermal dynamic phase
of the system. We have also checked the case of the reverse quench, where
we start from the state with conserved $U(N)$ and dipole symmetry.
In this case, the system always shows a single first-order
transition where both the symmetries are broken simultaneously.

The organization of rest of the paper is as follows.
In Sec.~\ref{eq1}, we introduce the boson model and discuss its
equilibrium phase diagram in the large-$N$ limit both at zero and
finite temperature. This is followed by Sec.~\ref{noneq1} where we
study non-equilibrium dynamics of the model for both quench and ramp
protocols. Finally, we discuss our main results and conclude in
Sec.~\ref{conc}.

\section{Equilibrium phases of bosons with dipole symmetry}
\label{eq1}

In this section, we first introduce the model that we
work with in Sec.~\ref{mod1} and chart out its equilibrium
properties using a large-$N$ approach. The corresponding large-$N$
solution is presented in Sec.~\ref{ln1}. The zero temperature phase
diagrams of the model based on the large-$N$ solution is discussed
in Sec.~\ref{zt1}.

\subsection{Model}
\label{mod1}

We consider a system of charged bosons in 3D,
whose long-wavelength low-energy thermal equilibrium properties at a
temperature $T$ are described by the Euclidean action
\begin{widetext}
\beq \label{action_scalar}
\begin{split}
S_E=\int d^3x
\int_0^\beta d\tau~&\left[\phi^{\ast}(x,\tau)(-\partial^2_{\tau}+m_0^2)\phi(x,\tau)+\lambda_4|\phi(x,\tau)|^4\right.\\ 
&\left.+\lambda\left\{
  ~\left(\phi^{\ast}(x,\tau)\right)^2 \left( \phi(x,\tau) \nabla^2
    \phi(x,\tau) -\vec{\nabla} \phi(x,\tau) \cdot \vec{\nabla} \phi(x,\tau)\right)+\text{h.c.}\right\} \right],
\end{split}
\eeq
\end{widetext}
where $\beta= 1/(k_B T)$ and $k_B$ is the Boltzmann
constant which shall be set to unity for the rest of this work.
This action, first proposed in Ref.~\onlinecite{fracton_gauge}, is
invariant under: (i) a global $U(1)$ transformation $\phi(x,\tau)
\rightarrow e^{i\alpha} \phi(x,\tau)$, which leads to
conservation of charge $\rho =\int d^3 x~ \rho(x)$ where
\begin{eqnarray}
\rho(x) &=& \int d^3 x~ \left(\partial_\tau \phi^\ast(x,\tau)
\phi(x,\tau)-\phi^\ast(x,\tau) \partial_\tau \phi(x,\tau) \right),
\nonumber\\ \label{charge}
\end{eqnarray}
and (ii) a dipole transformation $\phi(x,\tau) \rightarrow
e^{i\vec{\gamma}\cdot \vec{x}} \phi(x,\tau)$, which leads to the
conservation of the dipole moment 
\begin{eqnarray}
\vec d &=& \int d^3 x ~\vec{x} \rho(x). \label{dmom1}
\end{eqnarray}
Note that while the $U(1)$ transformation implements a global
symmetry, the dipole transformation has a status intermediate
between global and gauge symmetries. This is equivalent to model A
in Ref.~\onlinecite{largeN_fracton}.

The action Eq.~\ref{action_scalar} differs from usual interacting
complex scalar theories in the following ways: (i) there are no
spatial gradient terms in the quadratic part, i.e. there is no
microscopic single particle kinetic energy in the system. There is however a bare mass $m^2_0$ which gives the gap in the single-particle spectrum when interaction strengths are set to zero. (ii)
The presence of interaction terms ( $\sim \lambda$) which couple to
gradients of the fields and their powers. In addition, the system
has a usual $\phi^4$ type local interaction with a coupling strength
$\lambda_4$. The field theory with the dipole symmetry can in
principle have three different phases (a) where both $U(1)$ and
dipole symmetry is maintained (b) where dipole symmetry is broken
but $U(1)$ symmetry is present and (c) where both dipole and $U(1)$
symmetries are broken with the presence of a condensate. Note that
one cannot have a phase with broken $U(1)$ symmetry, where the
dipole symmetry is intact. The presence of the Goldstone modes would
ensure that the dipole symmetry is also broken in this case.

The standard complex scalar field theory in 3D has a quantum phase
transition as a function of the bare mass $m_0^2$ from a normal
phase to a $U(1)$ symmetry broken state with a condensate
characterized by $\langle \phi(x,\tau)\rangle =\sigma$. This naturally leads to the question on the nature of quantum
phase transitions, if any, for a scalar field theory with dipole
symmetry. We note that Ref.~\onlinecite{largeN_fracton} has
investigated this question for a theory with a single time
derivative (the Schr\"{o}dinger theory) and found that the dipole
symmetry is always broken in the mean field approximation, while the
system has a standard phase transition where the $U(1)$ symmetry is
broken. We will see that the theory with quadratic time derivative
exhibit a qualitatively different phase diagram \cite{fracton_gauge}.

In order to make analytic progress, we expand the ambit of the
problem to a field theory with $N$ flavours of charged bosons
\begin{widetext}
\bqa \label{action_scalar_N}
S_E&=&\int d^3x
\int_0^\beta d\tau~\left[\sum_a
  \phi^{\ast}_a~(x,\tau)(-\partial^2_{\tau}+m_0^2)\phi_a(x,\tau)+\frac{\lambda_4}{N}\left(\sum_a
    ~ |\phi_a(x,\tau)|^2\right)^2 \right.\\
  \nonumber & &\left. +\frac{\lambda}{N}
  ~\sum_{ab}~\left\{\phi^{\ast}_a(x,\tau)\phi^\ast_b(x,\tau) \left(\frac{1}{2} \left[\phi_a(x,\tau) \nabla^2
    \phi_b(x,\tau)+\phi_b(x,\tau) \nabla^2
    \phi_a(x,\tau)\right] -\vec{\nabla} \phi_a(x,\tau) \cdot \vec{\nabla} \phi_b(x,\tau)\right)+\text{h.c.}\right\} \right],
\eqa
\end{widetext}
where $a$ and $b$ denote the flavours of the bosons. The symmetry is now
enhanced to a $U(N)$ symmetry. It is useful to
Fourier transform the fields and work in the momentum-Matsubara
frequency coordinates,
\begin{widetext}
\beq\label{fracton_scalar_mom}
        S_E=\int Dk~\sum_a \phi^{\ast}_a(k)\left(\omega^2_n+m^2_0\right)\phi_a(k)
    +\int \prod_{i=1}^4Dk_i~\sum_{ab}~\phi^{\ast}_a(k_1)\phi^{\ast}_b(k_2)\phi_b(k_3)\phi_a(k_4)V(\{k_i\}),
\eeq
where $k=(\textbf{k},\omega_n)$,~$\omega_n=\frac{2\pi n}{\beta}$,~$\int
Dk=\frac{1}{\beta}\sum_{n}\int\frac{d^3k}{(2\pi)^3}$ and
\begin{eqnarray}
V(\{k_i\})&=& \left[\frac{\lambda_4}{N}+\frac{\lambda}{2N}
\left(|\textbf{k}_1-\textbf{k}_2|^2+|\textbf{k}_3-\textbf{k}_4|^2\right)\right] \times \delta(k_1+k_2-k_3-k_4). \label{Vk}
\end{eqnarray}
\end{widetext}
The continuum field theories require an ``UV" regularization scheme
for well defined answers for interacting correlators. This is
usually treated within standard renormalization schemes where
couplings run with cut-offs, leading to low-energy (``IR") physical
observables which are independent of the ``UV" regularization.
However, systems with dipole symmetry are known to exhibit strong ``UV-IR" mixing, e.g. low- energy properties like degeneracy of ground
states are sensitive to the ``UV" regularization. Here we will
regularize the theory on a cubic lattice of lattice spacing $a$ with
periodic boundary conditions, which amounts to replacing
$|\textbf{k}_i-\textbf{k}_j|^2$ in Eq.~\ref{Vk} by 
\begin{eqnarray}
\epsilon(\textbf{k}_i-\textbf{k}_j) &=& \frac{4}{a^2}
[\sin^2(k_i^x-k_j^x)a +\sin^2(k_i^y-k_j^y)a \nonumber\\
&& +\sin^2(k_i^z-k_j^z)a]. \label{disp1}
\end{eqnarray}
This in turn means that we work with the regularized interaction
\begin{eqnarray}
V_r(\{\textbf{k}_i\}) &=&
\left[\frac{\lambda_4}{N}+\frac{\lambda}{2N}
\left(\epsilon(\textbf{k}_1-\textbf{k}_2)+\epsilon(\textbf{k}_3-\textbf{k}_4|)\right)\right]\nonumber\\
&& \times \delta(k_1+k_2-k_3-k_4) \label{vertex_lattice}
\end{eqnarray}
The momenta are then restricted to the first Brillouin zone, $-\pi/a
\leq k^x,k^y,k^z \leq \pi/a$. In this lattice regularized theory, we
will now consider the phase diagram of the system in thermal
equilibrium in the large-$N$ limit.

\begin{figure*}[t]
    \includegraphics[scale=0.4]{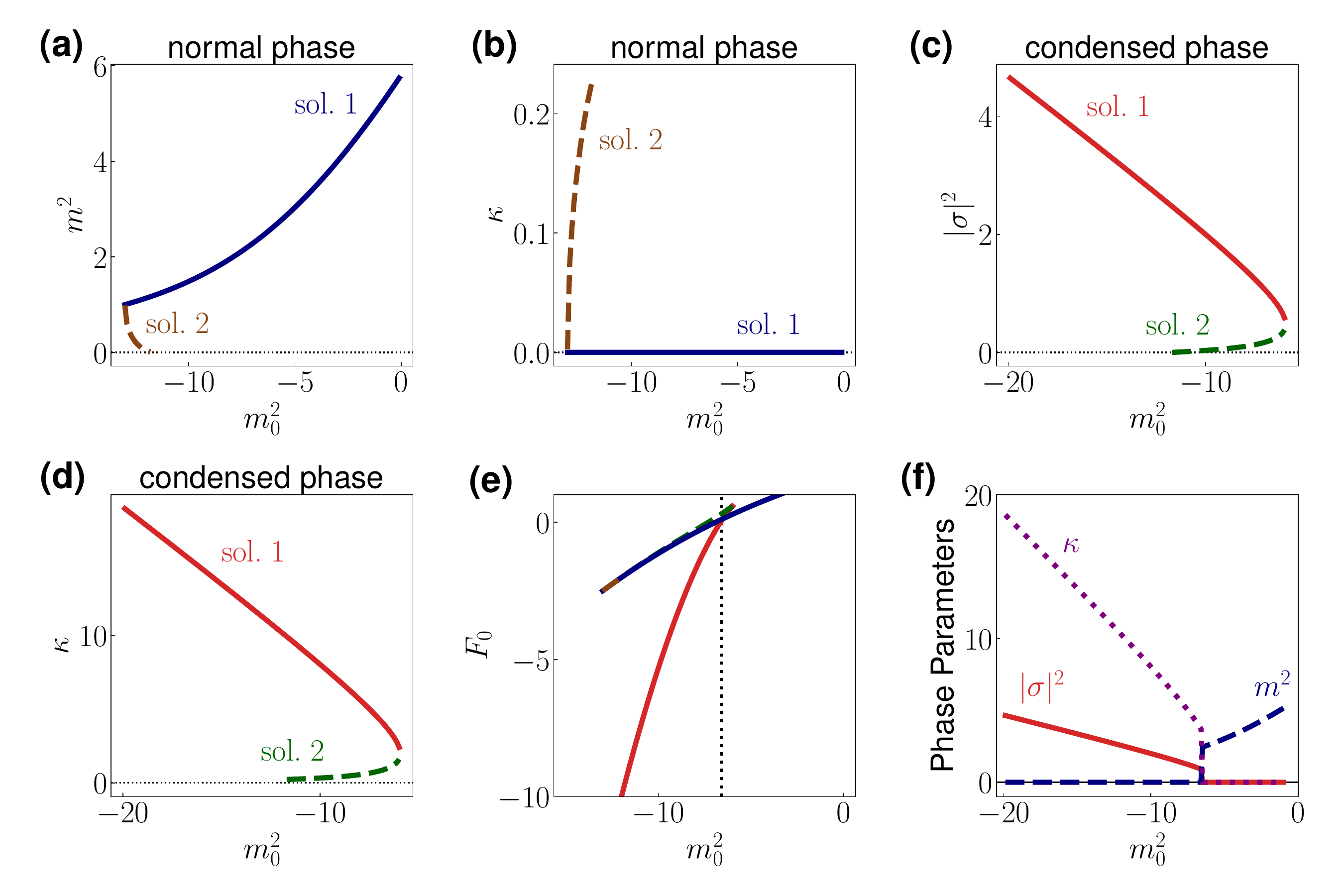}
    \caption{Equilibrium phases of dipole conserving bosons at zero temperature. (a), (b) Solutions of the self-consistent equations for normal phase (see Eq.~\ref{normeqs}) are presented as a function of $m^2_0$. One solution (shown in blue solid line) exist for $m_0^2>-13.0$. This solution conserves both $U(N)$ and dipole symmetry. The other solution (shown in brown dashed line) exists for $-13.0<m_0^2<-11.8$. This solution conserves $U(N)$ but breaks dipole symmetry. (c), (d) Solutions of the self-consistent equations for condensed phase (see Eq.~\ref{condeqs}) are presented as a function of $m^2_0$. One solution (shown in red solid line) exist for $m_0^2<-6.0$. The other solution (shown in green dashed line) exists for $-11.8<m_0^2<-6.0$. Both the solution breaks $U(N)$ and dipole symmetry. (e) Free energy corresponding to all the solutions presented are plotted as a function of $m^2_0$. The solution with the lowest free energy is the ground state. For, $m_0^2>-6.6$ the blue solution (normal phase) is the ground state and for $m_0^2<-6.6$ the red solution (condensed phase) is the ground state. (f) The phase parameters of the ground state as a function of $m^2_0$ is plotted. The blue dashed line, the purple dotted line and red solid line represent $m^2$, $\kappa$ and $|\sigma|^2$ respectively. For $m_0^2>-6.6$, the ground state is a normal phase with both $U(N)$ and dipole symmetry conserved ($|\sigma|^2=0,~\kappa=0$). At $m_0^2=-6.6$, both $\kappa$ and $|\sigma|^2$ jump to finite values while $m^2$ goes to zero. This represents a first- order transition to the condensed phase, which simultaneously breaks both $U(N)$ and dipole symmetries. For all the plots $\lambda=\lambda_4=2$ and energy is measured in the units of $\sqrt[3]{\lambda_4/2a^3}$.
    }
    \label{phasesT0}
\end{figure*}

\subsection{Large-$N$ solution}
\label{ln1}

In large-$N$ limit, bilinears of the fields, summed over flavour
indices, can be assigned an average value, i.e their fluctuations
are suppressed parametrically in powers of $1/N$. This mean field
decoupling leads to an effective Gaussian theory, where the single
particle Green's function $G$ is dressed by a self-energy $\Sigma$, which is
determined self-consistently. The details of the large-$N$
calculation are given in Appendix ~\ref{ApxlargeNsaddle} . The
single-particle Green's function is given by
\begin{subequations}\label{largeN_dyson}
    \begin{align}
    G(k)&=\frac{1}{\omega_n^2+\omega(\textbf{k})^2}+|\sigma|^2\beta\delta_{n0}(2\pi)^3\delta^3(\textbf{k})\label{largeN_dyson_a},
          ~\textrm{with}\\
    \omega(\textbf{k})^2&=m^2_0+2\int Dk'
    G(k')NV_r(\textbf{k},\textbf{k}';\textbf{k},\textbf{k}')\label{largeN_dyson_b},
\end{align}
\end{subequations}
 where $\sigma =\langle\phi_{a=1}\rangle/\sqrt{N}$ is
the condensate fraction. For $\sigma \neq 0$, the $U(N)$ symmetry is
broken down to $U(N-1)$, and Goldstone's theorem
forces the dispersion to be gapless at $\textbf{k}=0$. On the other
hand, if the dispersion is gapped, the system must be in its normal
state with $\sigma=0$ ($U(N)$ is intact as a symmetry of the phase).
This can be concisely written as a single condition, $\sigma
\omega(\textbf{k}=0)=0$.
Putting the expression for $G(k)$ (Eq.~\ref{largeN_dyson_a}) in the equation for $\omega(\textbf{k})$  (Eq.~\ref{largeN_dyson_b}) and performing the Matsubara sum we get the self-consistent equation,
\beq\label{largeN_selfcon}
\begin{split}
    \omega(\textbf{k})^2&=m^2_0+2|\sigma|^2NV_r(\textbf{k},0;\textbf{k},0)\\&+2\int\frac{d^3k'}{(2\pi)^3}NV_r(\textbf{k},\textbf{k}';\textbf{k},\textbf{k}')\frac{\coth\left(\frac{\omega(\textbf{k}')}{2T}\right)}{2\omega(\textbf{k}')}
\end{split}
\eeq  Using Eq.~\ref{largeN_selfcon}, one can easily show that the dispersion can be parametrized as
\beq
 \omega(\textbf{k})^2=m^2+\kappa\epsilon(\textbf{k}).
\eeq
Putting the expressions for $NV_r(\textbf{k},\textbf{k}';\textbf{k},\textbf{k}')$ and $\omega(\textbf{k})$ in Eq.~\ref{largeN_selfcon} we get self-consistent equations for the parameters. For the normal phase, the equations are
\begin{subequations}\label{normeqs}
    \begin{align}
    m^2&=m_0^2+2\lambda_4I_1(m^2,\kappa)+2\lambda I_2(m^2,\kappa), \label{normeq1}\\
    \kappa&=2\lambda \left[I_1(m^2,\kappa)-(a^2/6)I_2(m^2,\kappa)\right],\label{normeq2}
    \end{align}
\end{subequations}
and for the condensed phase, the equations are
\begin{subequations}\label{condeqs}
    \begin{align}
    &\frac{\lambda_4}{\lambda}\kappa+2[\lambda +\lambda_4(a^2/6)]I_2(0,\kappa)+m_0^2=0 , \label{condeq1}\\
    &|\sigma|^2=\frac{\kappa}{2\lambda}-I_1(0,\kappa)+(a^2/6)I_2(0,\kappa),\label{condeq2}
    \end{align}
\end{subequations}
where
\begin{subequations}\label{integs}
\begin{align}
    I_1(m^2,\kappa)&=\int\frac{d^3k}{(2\pi)^3}\frac{\coth\left(\frac{\omega(\textbf{k})}{2T}\right)}{2\omega(\textbf{k})},\label{integ1}\\
    I_2(m^2,\kappa)&=\int\frac{d^3k}{(2\pi)^3}\frac{\epsilon(\textbf{k})\coth\left(\frac{\omega(\textbf{k})}{2T}\right)}{2\omega(\textbf{k})}\label{integ2}.
\end{align}
\end{subequations}
We are now in a position to discuss the significance of the
parameters $m^2$, $\kappa$ and $\sigma$. $m^2$ is the effective mass
of the interacting particles, while a non-zero $\kappa$ denotes the
presence of an effective kinetic energy or dispersion.  If $\sigma=0$, i.e. if the $U(N)$ symmetry is not broken,
$m^2$ is finite; in contrast, $\sigma \ne 0$ indicates the
presence of a condensed phase with broken $U(N)$ symmetry ($U(N)
\rightarrow U(N-1)$)and
$m^2=0$. Moreover, if $\kappa \neq 0$, the dipole symmetry is
broken, while $\kappa=0$ denotes a system with intact dipole
symmetry. For $\sigma\neq 0$, one would generically expect $\kappa
\neq 0$ as the dipole symmetry is also broken in this case. Thus
one can determine the symmetry of the phase simply by tracking the
behaviour of $\sigma$ and $\kappa$.

\subsection{Phase Diagram at $T=0$}
\label{zt1}

We now consider the phase diagram of the system at $T=0$ as a
function of the bare mass $m_0^2$. We set $a=1$,
$\coth(\omega(\textbf{k})/2T)=1$, and fix the couplings $\lambda=2$
and $\lambda_4=2$.

Let us consider the solutions to the normal state ($\sigma=0$) self
consistent equations Eq.~\ref{normeqs}. The solutions for $m^2$ are
plotted in Fig.~\ref{phasesT0}(a). We find that solutions exist for
$m_0^2 >-13.0$. The blue solid line corresponds to a solution where
$m^2$ increases with increasing $m_0^2$. This corresponds to a
solution with $\kappa=0$, i.e. both $U(N)$ and dipole symmetry are
intact for this solution. For a small range of $-13.0 < m_0^2 <
-11.8$, another solution exists, which is shown with dashed brown
line. For this solution $\kappa \neq 0$ (see Fig.~\ref{phasesT0}(b)
), i.e. the dipole symmetry is broken although $U(N)$ is preserved.
The mass $m^2$ decreases with increasing $m_0^2$ for this solution.

We now consider the solutions of the self-consistent
equations (Eq.~\ref{condeqs}) in the condensed phase ($m^2=0$,
$\sigma \neq 0$). As shown in Fig.~\ref{phasesT0}(c), one solution,
shown in solid red, increases with decreasing $m_0^2$, and exists
for $m_0^2<-6.0$. For $-11.8 < m_0^2 <-6.0$, another solution (shown
in dashed green), which decreases with decreasing $m_0^2$ exists.
Both solutions have finite $\kappa$ (Fig.~\ref{phasesT0}(d)) and
break both $U(N)$ and dipole symmetry.

The existence of multiple solutions to the self-consistent equations
in overlapping region of phase space indicates the
occurrence of metastable states. In such a case the system is likely to
undergo a first-order phase transition based on the energetics of the
system. To consider this, we compute the on-shell action (free energy)
of the different solutions and plot it as a function of $m_0^2$ in Fig.~\ref{phasesT0}(e). We
see that the lowest energy solution is the blue solution in the
normal state  for $m_0^2> -6.6$ and the red solution in the condensed
phase for $m_0^2<-6.6$. Thus, the system transitions from a phase with
both $U(N)$ and dipole symmetry to a phase where both symmetries are
broken through a first-order transition. The possible normal phase
with broken dipole symmetry does not show up in this phase diagram. The parameters of the lowest
energy solution are plotted as a function of $m_0^2$ in
Fig.~\ref{phasesT0}(f). $m^2$ , $\kappa$ and $\sigma$ all show a finite
jump at $m_0^2=-6.6$, which is the phase transition point.

We note that we have also calculated the phase diagram of the system
for different values of $\lambda/\lambda_4$. As long as both couplings
are finite we qualitatively find the same phase diagram, although the
location of the first-order transition points change and the jump
discontinuities increase with increasing $\lambda/\lambda_4$. Thus the gradient terms in the interaction drive the system towards stronger first-order transition. See
Appendix~\ref{ApxPhaseSol} for more details.

One can also solve the self-consistent equations at finite temperature
to obtain the thermal phase diagram of the system by tracking $m^2(T)$, $\sigma^2(T)$ and
$\kappa(T)$ (see Appendix~\ref{ApxPhaseTh} for details). We find that the thermal phase diagram also consists of a
fully symmetry conserved phase and a fully symmetry broken phase,
separated by a direct first-order phase transition, similar to the
$T=0$ phase diagram. The critical mass for the transition shifts to
the left as temperature is increased, producing the phase diagram in Fig.~\ref{quenchcond}(e).

\section{Non Equilibrium Dynamics: Quench and Ramp Protocols}
\label{noneq1}

 In this section, we consider non-equilibrium dynamics
of this system of charged bosons with dipole symmetry (in the large
$N$ limit). We will be interested in the dynamics of the system
after quantum quenches where parameters of the system are changed
instantaneously, as well as after a ramp involving change of system
parameters at a finite rate. We use a large-$N$ Keldysh technique
for this purpose which we introduce in Sec.\ \ref{kf1}. This is
followed by a discussion of quench and ramp dynamics of the system
starting from the condensed phase in Sec.\ \ref{cpdyn1} and from the
normal phase in Sec.\ \ref{npdyn1}. In Sec.~\ref{steady_osc}, we explain the nature and origin of the large oscillations seen in the steady state for some values of the quench parameters.

\subsection{Keldysh formalism}
\label{kf1}

 We work with the Schwinger-Keldysh formalism, where
the non-equilibrium dynamics is described in terms of two fields at
each space-time point \cite{kamenevbook,kamenevarticle}. This formalism has been
used extensively for treating non-equilibrium quantum field theory
in several contexts \cite{cgs,cgs_memory,rajdeep1,realscalar,others1}, including field theories
in their large-$N$ limit
\cite{sdas1,sondhi1,sondhi2,schiro1,vikram1,chang1,diptarka1}.

Using the standard classical  and ``quantum''
fields, the Schwinger-Keldysh action is given by \cite{kamenevarticle,realscalar}
\begin{widetext}
\beq
S= \int ~dt \int \frac{d^3k}{(2\pi)^3} \left(\phi_{cl}^\ast(\textbf{k},t),
  \phi_{q}^\ast(\textbf{k},t)\right)\left(\begin{array}{cc}
 0 & 2[-\partial_t^2 -m_0^2(t)]\\
 2[-\partial_t^2 -m_0^2(t)] & -\Sigma^K\delta(t)\end{array}\right) \left(\begin{array}{c}
\phi_{cl}(\textbf{k},t),\\
\phi_{q}(\textbf{k},t)\end{array}\right) +S_{int}
\eeq
where
\bqa
S_{int}&=&-\int ~dt \int \prod_{i=1}^4 \frac{d^3 k_i}{(2\pi)^3}~\delta(\textbf{k}_1+\textbf{k}_2-\textbf{k}_3-\textbf{k}_4)~
V_r(\{\textbf{k}_i\})\\
\nonumber & & 4\left[ \phi^\ast_{cl}(\textbf{k}_1,t)
  \phi^\ast_{cl}(\textbf{k}_2,t) \phi_{cl}(\textbf{k}_3,t)
  \phi_{q}(\textbf{k}_4,t) +\phi^\ast_{q}(\textbf{k}_1,t)
  \phi^\ast_{q}(\textbf{k}_2,t) \phi_{q}(\textbf{k}_3,t)
  \phi_{cl}(\textbf{k}_4,t) +~\textrm{h.c.}\right]
\eqa \end{widetext}
and we have chosen the bare mass
$m_0^2$ to be time dependent according to a quench or a ramp protocol. The
former involves a sudden change of $m_0$ from $(m_{0})_{in}$ to
$(m_{0})_f$, while the latter implements a gradual ramp with a rate
$\tau^{-1}$: $m^2(t)= (m_0)_{in}^2
+((m_0)_f^2-(m_0)_{in}^2)f(t/\tau)$, with $f(0)=0$ and $f(\infty)=1$.
A specific example, $f(x) = \tanh x$, shall be studied in details in
this work.

In the Schwinger-Keldysh formalism, there are two independent
one-particle correlators, the retarded Green's function $G^R(\textbf{k},t,t')
=-i \langle \phi_{cl}(\textbf{k},t) \phi^\ast_q
(\textbf{k},t')\rangle$, which controls the evolution of the system
and the Keldysh Green's function $G^K(\textbf{k},t,t')
=-i \langle \phi_{cl}(\textbf{k},t) \phi^\ast_{cl}
(\textbf{k},t')\rangle$, which contains information about the time
evolving one-particle distribution function in the system.

In the large-$N$ limit, the effect of the interactions can be
absorbed into a retarded self-energy (see Appendix~\ref{ApxDynEq} for
details)
\begin{eqnarray}
\Sigma^R(\textbf{k},t) &=& \mathbf{i} \int\frac{d^3k'}{(2\pi)^3}
4V_r(\textbf{k},\textbf{k}',\textbf{k},\textbf{k}')
G^K(\textbf{k}',t,t)\nonumber\\
&& +\frac{\mathbf{i}}{(2\pi)^3}4V_r(\textbf{k},0,\textbf{k},0)
G^K(0;t,t). \label{geq1}
\end{eqnarray}
In this case, the retarded Green's function can be obtained by
solving 
\beq 
2\left[-\partial_t^2 -m^2(t) -\kappa(t)
\epsilon(\textbf{k})\right] G^R(\textbf{k},t,t')= \delta(t-t'),
\label{meq1} 
\eeq 
where
\begin{widetext}
\bseq\label{dynparams}
\begin{align}
	m^2(t)&=m_0^2(t)+2\lambda_4|\sigma|^2(t)+2\lambda_4\int\frac{d^3k}{(2\pi)^3}iG^K(\textbf{k};t,t)+2\lambda\int\frac{d^3k}{(2\pi)^3}\epsilon(\textbf{k})iG^K(\textbf{k};t,t)\\
	\kappa(t)&=2\lambda|\sigma|^2(t)+2\lambda\int\frac{d^3k}{(2\pi)^3}iG^K(\textbf{k};t,t)-2\lambda(a^2/6)\int\frac{d^3k}{(2\pi)^3}\epsilon(\textbf{k})iG^K(\textbf{k};t,t)\\
	|\sigma|^2(t)&=\frac{1}{(2\pi)^3}iG^K(0;t,t), 
\end{align}
\eseq
To complete the set of self-consistent equations, we need to write
$G^K$ in terms of $G^R$. This is done through the Dyson equation (see
Appendix~\ref{ApxDynEq} and Ref.~\onlinecite{realscalar} for
details),
\beq
	G^K(\textbf{k};t,t)=
	[2\omega_{in}(\textbf{k})]^2G^K(\textbf{k};0,0)\Big
	\{|G^R(\textbf{k};t,0)|^2+\frac{1}{[\omega_{in}(\textbf{k})]^2}|\bar{G}^{R}(\textbf{k};t,0)|^2\Big
	\}, \label{sc1}
\eeq
\end{widetext}
where $\bar{G}^R(t,t')=\partial_{t'} G^R(t,t')$ and we have assumed
that the system was initialized to Fock states corresponding to
initial dispersion $\omega_{in}(\textbf{k})$. For both the quench and the ramp dynamics $\omega_{in}(\textbf{k})$ is the $T=0$ equilibrium spectrum at $(m^2_0)_{in}$. Eqs.\
\ref{geq1}, \ref{meq1}, \ref{dynparams} and \ref{sc1} therefore
constitutes a closed set of coupled equations which provides a
concrete prescription to study the non-equilibrium dynamics of this
system in the large-$N$ approximation.

\subsection{Dynamics starting from the condensed phase}
\label{cpdyn1}

\begin{figure*}[t]
	\includegraphics[scale=0.38]{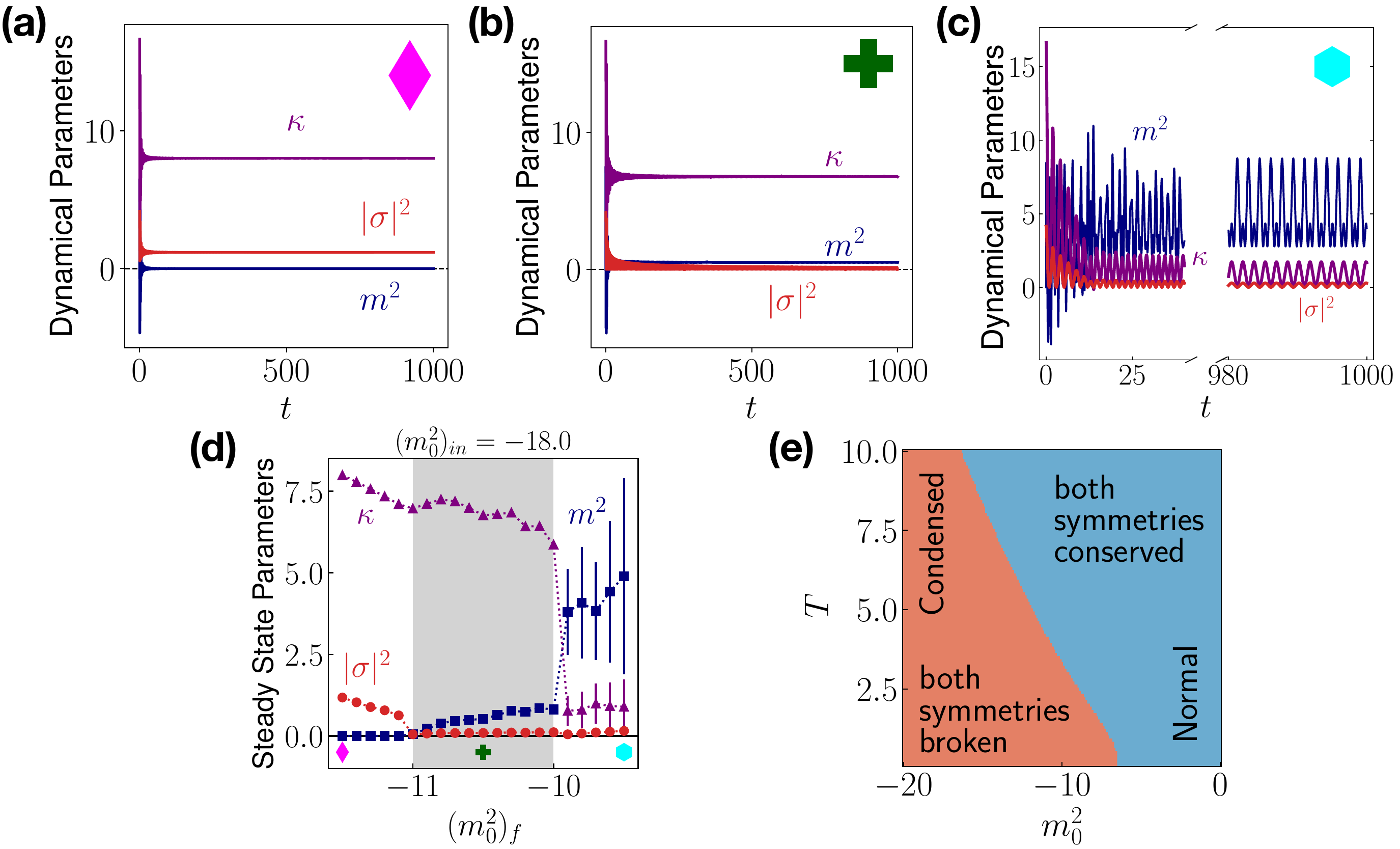}
	\caption{Dynamics of parameters after a sudden quench of $m^2_0$ starting from ground state in condensed phase with $(m^2_0)_{in}=-18.0$ and $\lambda=\lambda_4=2$. The initial values of the parameters are $(|\sigma|^2)_{in}=4.15$, $(\kappa)_{in}=16.63$ and $(m^2)_{in}=0$. $(m^2_0)_{f}$ denotes the final value of $m^2_0$ after the quench. Here the time is measured in the units of $\sqrt[3]{2a^3/\lambda_4}$.
		(a) Dynamics of the parameters for $(m^2_0)_{f}=-11.5$ is shown. At long times, $m^2=0$ while $\kappa$ and $|\sigma|^2$ remain finite. This represents a condensed phase.
		(b) Dynamics of the parameters for $(m^2_0)_{f}=-10.5$ is shown. At long times, $m^2$ gains finite value while $|\sigma|^2$ shows small oscillation with its lowest value touching 0 and  $\kappa$ remains finite. This represents a phase where dipole symmetry is broken but $U(N)$ is restored. This is a dynamical phase which is absent at equilibrium.
		(c) Dynamics of the phase parameters for $(m^2_0)_{f}=-9.5$ is shown. The x-axis is broken to accommodate both the initial transients and steady state dynamics. At long times, $m^2$ fluctuates largely with a finite time-averaged value. It stays well above 0. $\kappa$ and $|\sigma|^2$ fluctuates with their lowest value very close to 0. This represents a normal phase.
		(d) Steady state time-averaged values (averaged for $t=500-1000$) of the dynamical parameters are shown as a function of $(m^2_0)_f$. The error-bars represent amplitude of the oscillations at long times. We see two transitions separating three kinds of phases. At $(m^2_0)_{f}=-11.0$, $|\sigma|^2$ smoothly goes to 0 (its lowest value touches 0) and $m^2$ attains finite value while $\kappa$ remains finite. At $(m^2_0)_{f}=-9.9$, $\kappa$ falls sharply and $m^2$ rises further. So, for $(m_0^2)_f<-11.0$, we have condensed phase with charge and dipole symmetry broken. For $(m_0^2)_f>-9.9$, we have the normal phase with conserved charge and dipole. Between these two transitions, charge is conserved but dipole is not. This is a dynamical phase which is absent in equilibrium. So, $U(N)$ and dipole symmetries get restored at different points unlike the case in equilibrium where this happens in a single transition. The positions of $(m^2_0)_{f}$ values for which the dynamics of the dynamical parameters are shown in (a), (b) and (c) are indicated by a magenta diamond, a green `+' and a cyan hexagon respectively.
		(e)Thermal phase diagram of bosons with dipole symmetry in the $m_0^2-T$ plane. The light red shaded area hosts condensed phase ($m^2=0,\kappa>0,|\sigma|^2>0$) where $U(N)$ and dipole symmetries both are broken. The light blue shaded area is for the normal phase ($m^2>0,\kappa=0,|\sigma|^2=0$) where both the $U(N)$ and dipole symmetries are conserved. The boundary between these shaded areas indicate a first- order transition. The critical $m^2_0$ decreases with increasing temperature. Note that the intermediate dynamical phase with conserved $U(N)$ but broken dipole symmetry is absent in the thermal phase diagram.}
	\label{quenchcond}
\end{figure*}

We consider the situation where the system of bosons is initialized
to the interacting ground state corresponding to the couplings
$\lambda$, $\lambda_4$ and $\left(m_0^2\right)_{in}$. The parameters
are so chosen that the system is in the symmetry broken condensed
phase with $m^2_{in}=0$, $|\sigma|^2_{in} \neq 0$ and $\kappa_{in}
\neq 0$. This implies that $\omega_{in}^2(\textbf{k}) =
\kappa_{in} \epsilon(\textbf{k})$ for our problem. For most of the
calculations, we will use $\lambda=\lambda_4=2$ and
$\left(m_0^2\right)_{in}=-18.0$, unless otherwise specified.

The bare mass of the system is then instantaneously changed from
$\left(m_0^2\right)_{in}$ to $\left(m_0^2\right)_{f}$, and the
system is allowed to evolve under a closed system dynamics governed by Eqs.~\ref{geq1}-\ref{sc1}. The instantaneous
state of the system can be characterized by $m^2(t)$ and $\kappa(t)$
which gives the instantaneous dispersion, $|\sigma|^2(t)$, which
denotes the condensate density and $G^K(\textbf{k},t,t)$, which
keeps track of the distribution function in the system. The result of such a post-quench evolution is shown in Fig.\
\ref{quenchcond}.

Fig.~\ref{quenchcond} (a) shows the time evolution of the parameters
$m^2(t)$, $\kappa(t)$ and $|\sigma|^2(t)$, after a quench from
$\left(m_0^2\right)_{in}=-18.0$ to $\left(m_0^2\right)_{f}=-11.5$,
i.e. the final parameters also correspond to the equilibrium system
in the condensed phase. In this case, it is clear that the
parameters settle into a quasi-stationary state (parameters oscillate
around an average value, although the amplitude of the oscillations
are too small to be seen in the figure), where the average of $m^2
=0$, while $|\sigma|^2$ and $\kappa$ settle into a pattern with
non-zero average values. In this case, the steady state has broken
both the $U(N)$ ($U(N) \rightarrow U(N-1)$) and the dipole symmetry.

Fig.~\ref{quenchcond} (b) shows the evolution of the parameters
after a quench from $\left(m_0^2\right)_{in}=-18.0$ to
$\left(m_0^2\right)_{f}=-10.5$. The equilibrium state at the final
parameters would still correspond to a condensed phase. Once again
the parameters settle down to quasi-stationary states, where
$|\sigma|^2$ oscillates with very small amplitude and its lower end
touches $0$, while $m^2$ has a small but finite value, i.e. this is
a normal state where the $U(N)$ symmetry is restored. However we see
that the average value of $\kappa$ is finite and the state thus
breaks dipole symmetry. We thus recover the dipole symmetry broken
normal state, which is absent in the equilibrium phase diagram, as a
dynamical phase corresponding to the steady state of the system
after a quantum quench.

Fig.~\ref{quenchcond}(c) shows the evolution of the parameters after
a quench from $\left(m_0^2\right)_{in}=-18.0$ to
$\left(m_0^2\right)_{f}=-9.5$. The equilibrium state at the final
parameters would still correspond to a condensed phase. We see that
the steady (quasi-stationary) state has $|\sigma|^2=0$ and a finite
gap $m^2$. $m^2$ has large amplitude oscillations around the average
value, but the lowest value of $m^2$ remains positive. The curvature
$\kappa$ shows large oscillations around a small value, such that
the lowest value of $\kappa$ reaches $0$. We identify this as the
phase where both dipole and $U(N)$ symmetry are restored.

\begin{figure}[h!]
	\includegraphics[scale=0.30]{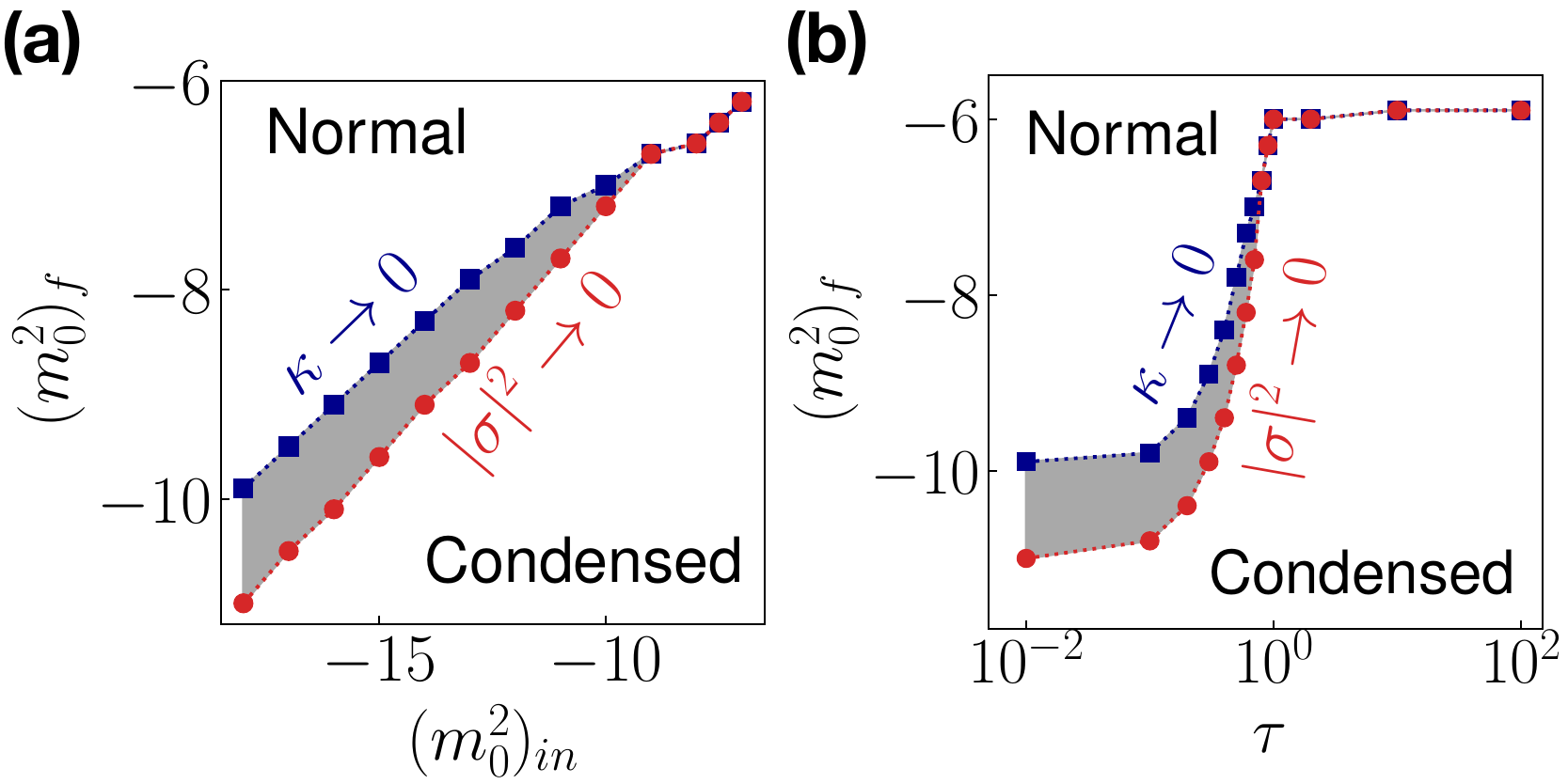}
	\caption{Dynamical phase diagram for non-equilibrium dynamics of bosons with dipole symmetry: (a) The steady state phases after a quantum quench from $(m^2_0)_{in}$ to $(m^2_0)_f$. The normal phase has conserved $U(N)$ and dipole symmetry, while both symmetries are broken in the condensed phase. The shaded region shows the dynamical phase with $U(N)$ symmetry, where dipole conservation is broken. The phase boundaries are obtained by considering  critical quench parameters where $|\sigma|^2\rightarrow0$ (red points) and $\kappa\rightarrow0$ (blue points) in the steady state. Note that for $(m^2_0)_{in}>-9.0$, the system has direct first-order transition from condensed to normal phase. The intermediate phase exists for $(m^2_0)_{in}<-9.0$ and occupies larger phase-space as $(m^2_0)_{in}$ is reduced. (b) The steady state phases after a ramp of $m^2_0$ from $(m^2_0)_{in}=-18.0$ to $(m^2_0)_{f}$ with a ramp timescale $\tau$  (in the $\tau-(m^2_0)_f$ plane). In the normal (condensed) phase both $U(N)$ and dipole symmetries are present (broken), while the shaded region shows the intermediate phase, where $U(N)$ symmetry is conserved but dipole symmetry is broken. Note that the intermediate phase vanished beyond $\tau=0.8$, and the system shows a first-order transition from condensed to normal phase beyond this value of $\tau$ (similar to the equilibrium phase diagram). Here $\lambda=\lambda_4=2$ for both the plots and $\tau$ is measured in the units of $\sqrt[3]{2a^3/\lambda_4}$.
	}
	\label{quenchinterim}
\end{figure}

We now consider the average long time value of the steady state
parameters as a function of the final quenched mass
$\left(m_0^2\right)_{f}$. This average is calculated by averaging
the values of the parameters over a time range from $500$ to $1000$ to get
rid of the effects of oscillations. Here the time is measured in the units of $\sqrt[3]{2a^3/\lambda_4}$. We have checked that changing
the range of this averaging does not change the answers
qualitatively, and changes them quantitatively by at most $2\%$.

These average values are plotted in Fig.~\ref{quenchcond}(d) with
the amplitude of the oscillations setting the errorbar shown in the
figure. We find that for $\left(m_0^2\right)_{f} < -11.0$, the
condensate density and the curvature is finite, while the gap is
zero. At $\left(m_0^2\right)_{f} = -11.0$, the condensate density
continuously goes to zero through a second order
dynamical transition; the gap also starts becoming finite just
beyond this point. Note that the curvature $\kappa$ is finite and
shows a smooth behaviour around this transition point.
Thus, one expect this transition to be a second order transition,
across which $U(N)$ symmetry is broken down to $U(N-1)$. As the final mass $(m_0)_f$
is increased, at $\left(m_0^2\right)_{f} = -9.9$, the curvature
shows a rapid decrease while the gap jumps to a
finite value; this constitutes a clear non-analytic behaviour. The
fluctuations in the gap and the curvature, as measured by the
oscillation amplitudes grow significantly beyond this jump (In
Fig.~\ref{quenchcond}(d), the error bars are smaller than the symbol
size for $\left(m_0^2\right)_{f} < -9.9$). Note that the average
curvature does not immediately go to zero, but its lowest value goes
below it eventually. So, we consider the sharp change in $\kappa$ and $m^2$ as the indication of the transition point. We thus have a second transition where the dipole symmetry
is restored. We use the point where the curvature and gap shows a
jump discontinuity to define the first-order transition which
restores dipole symmetry. Both these transitions occur to the left
of the equilibrium transition at $m_0^2=-6.6$. Thus the single first
order transition in equilibrium is split into a continuous
transition and a first-order transition in the dynamic phase diagram
of the system.

Till now, we have discussed the phase diagram of the steady state
after a quench from $\left(m_0^2\right)_{in}=-18.0$ to different
values of $\left(m_0^2\right)_{f}$. However, one can ask: does the
presence/size of the intermediate phase depend on the initial
mass from where we are starting our quench? To see
this, we first vary the initial $\left(m_0^2\right)_{in}$ and plot the
critical quenched $\left(m_0^2\right)_{f}$ where the jump occurs and
where the condensate vanishes. The two critical quench parameters (
final bare mass where the transitions take place) are
plotted as a function of $\left(m_0^2\right)_{in}$ in
Fig.~\ref{quenchinterim}(a). The intermediate phase exists when these two values are
different as shown by the shaded region. We see that the intermediate phase exists for
$\left(m_0^2\right)_{in} <-9.0$. For $\left(m_0^2\right)_{in} >-9.0$,
which is still in the condensed phase, the system directly jumps from
the symmetry broken phase to the phase where both symmetries are
restored. Note that while different initial quench masses lead to
different amount of energy dumped in the system, this is not the sole
parameter governing the phase diagram (e.g. the energy dumped at the
critical final mass varies quite a lot with changes in the initial mass).

There is a different way to control the amount of energy dumped due to
change in parameters of the system. Instead of an instantaneous
quench, one can change $m_0^2$ smoothly in time through
\beq\label{ramptanh}
m^2_0(t)=\left(m_0^2\right)_{in}+\left[\left(m_0^2\right)_{f}-\left(m_0^2\right)_{in}\right]\tanh(t/\tau),
\eeq
where $\tau$ is the ramp timescale. $\tau \rightarrow 0$ leads to an
instantaneous quench, while $\tau \rightarrow \infty$ depicts an
adiabatic change of the parameter. We track the dynamics of the system
undergoing this ramp at large
times ($t \gg \tau$) and obtain steady average values of $m^2$ and
$\kappa$. In Fig.~\ref{quenchinterim}(b), the
two critical  $\left(m_0^2\right)_{f}$s,  where $\sigma^2$ vanishes and
where $\kappa$ and $m^2$ jumps, are plotted as a function of the ramp
timescale. The phase space for the intermediate dipole symmetry broken
normal state (shown by the shaded region) shrinks
with increasing $\tau$ and vanishes beyond a finite value of $\tau=0.8$. The system thus undergoes a dynamical phase transition as a
function of the ramp time, with the intermediate phase absent in the
adiabatic limit. In the adiabatic limit, the system follows the
condensed phase and directly shows a first-order transition to a
normal state with conserved dipole symmetry. We thus see that the ramp
rate is a crucial parameter which determines the phase diagram of the system.

Finally, we consider what is likely to happen if we treat the dynamics
beyond the large-$N$ approximation. The sub-leading terms in the action
would introduce scattering between the momentum modes, effectively
introducing dissipation for the one particle dynamics. In this case,
the intermediate phase will disappear over a longer timescale and the
system will thermalize at the longest time-scales. However since the
time-scale is parametrically larger than the timescales involved in
the large-$N$ dynamics, one would expect the intermediate phase to
show up as a long-lived prethermal state of the system.

\subsection{Dynamics starting from the normal phase}
\label{npdyn1}
\begin{figure*}[t]
	\includegraphics[scale=0.4]{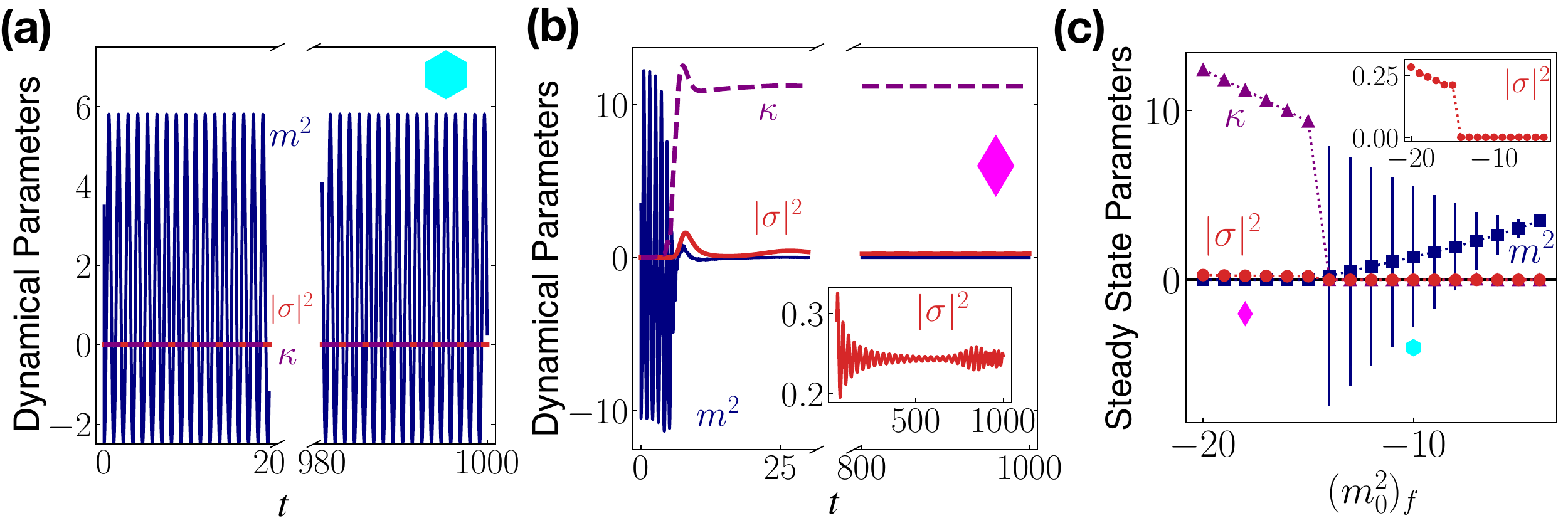}
	\caption{Dynamics of parameters after a sudden quench of $m^2_0$ starting from ground state in normal phase with $(m^2_0)_{in}=-4.0$ and $\lambda=\lambda_4=2$. Initial values of the phase parameters are $(m^2)_{in}=3.49$, $(\kappa)_{in}=0.0$ and $(|\sigma|^2)_{in}=10^{-4}$. $(m^2_0)_{f}$ denotes the final value of $m^2_0$ after the quench. The time is measured in the units of $\sqrt[3]{2a^3/\lambda_4}$.
		(a) Time dependence of the parameters for $(m^2_0)_{f}=-10.0$. $m^2$ oscillates around a finite value, while $\kappa$ and $|\sigma|^2$ vanish. This represents a normal phase where both the symmetries are conserved.
		(b) Time evolution of the parameters for $(m^2_0)_{f}=-18.0$. At long times, $m^2$ goes to zero and $\kappa$ and $|\sigma|^2$ (as shown in the inset) gain finite values. This indicates that the system goes to the condensed phase at long times. In (a) and (b) the x-axis is broken to accommodate both the initial transients and steady state dynamics.
		(c) Long time average values (averaged for $t=500-1000$) of the dynamical parameters are shown as a function of $(m^2_0)_f$. The error-bars represent amplitudes of the oscillations a long times. As $(m^2_0)_f$ is decreased, $m^2$ drops but its fluctuation increases until $(m^2_0)_{f}=-14.9$ where both its average value and fluctuation go to zero. Immediately after this point, $\kappa$ jumps to finite value. At the same point the condensate $|\sigma|^2$ also attains finite value, which is significantly higher than the initial seed value of $10^{-4}$, as shown in the inset. This indicates a first-order transition to the condensed phase. The final masses for which dynamics of the parameters are shown in (a) and (b) are indicated by a cyan hexagon and a magenta diamond respectively in (c).}
	\label{quenchnorm}
\end{figure*}

It is well-known that systems undergoing first-order
transitions typically show hysteresis, and so one would normally
expect that the dynamics of the system starting from the condensed
phase will not be the same as the dynamics of the system starting
from the normal phase. We will now focus on the dynamics of this
system starting from a ground state in the normal phase, which
preserves both the $U(N)$ and the dipole symmetry.

We consider the dynamics of the system after a sudden quantum quench. The system is initialized in
the ground state of normal phase with $\left(m_0^2\right)_{in}=-4.0$
and $m_0^2$ is suddenly quenched to a more negative
value $\left(m_0^2\right)_{f}$. At the initial point, $\kappa_{in}=0$ and
$m^2_{in}=3.49$. To allow for the possibility that the dynamics can
lead to symmetry broken steady state, we start with a tiny seed value
for the condensate with $(|\sigma|^2)_{in}=10^{-4}$. One can think of
this as a seed value which may grow, shrink or remain the same under
the dynamics. The system is then allowed to evolve to a steady state
under its own dynamics, where the parameters oscillate around an
average value. In Fig.~\ref{quenchnorm}(a), we show the time dependence of the parameters $|\sigma|^2(t)$, $\kappa(t)$ and $m^2(t)$ after a quench to $(m^2_0)_f=-10$. We find that $\kappa$ and $|\sigma|^2$ vanishes in the steady state, while $m^2$ oscillates around a finite value, i.e. the steady state has $U(N)$ and dipole symmetries. In Fig.~\ref{quenchnorm}(b), we show the same parameters after a quench to $(m^2_0)_f=-18$. In this case, the steady state has $m^2=0$ and finite $\kappa$. We see from the inset that the condensate fraction $|\sigma|^2$ is non-zero in this steady state. The system is thus in a state with broken $U(N)$ and dipole symmetries. In Fig.~\ref{quenchnorm}(c), we present the steady
state average values (averaged for $t=500-1000$ in the units of $\sqrt[3]{2a^3/\lambda_4}$) of the
parameters as a function of $\left(m_0^2\right)_{f}$. The error bar represents the amplitude of oscillation around the average value. 
As $\left(m_0^2\right)_{f}$ is decreased, the gap $m^2$ drops but its
fluctuation increases until a point where both its average value and
fluctuation go to zero at  $\left(m_0^2\right)_{f}= -14.9$. Immediately after this, the curvature
$\kappa$ becomes finite with very small fluctuations (not visible on
the scale of the figure). At the same point, the condensate
$|\sigma|^2$ also attains small finite value, as seen in the inset of Fig.~\ref{quenchnorm}(c). Note that this value is much larger than the
seed value used, which indicates the instability towards a broken
symmetry phase. Note that both $\kappa$ and $|\sigma|^2$ shows
a jump discontinuity at the same point, thus the steady state under
the reverse dynamics from
the normal phase shows a single first-order transition, where both the
$U(N)$ and the dipole symmetry are broken simultaneously, similar to the equilibrium phase diagram. 
We have checked that changing the seed value for $|\sigma|^2$ to $10^{-6}$ changes the location of the transition slightly ($(m^2_0)_f$ changes from $-14.9$ to $-15.2$). However, the value of $|\sigma|^2$ in the steady state in the symmetry broken phase depends strongly on the seed value, as one would expect for an instability.

\subsection{Oscillations in the Steady State\label{steady_osc}}
\begin{figure}[!t]
	\centering
	\includegraphics[scale=0.31]{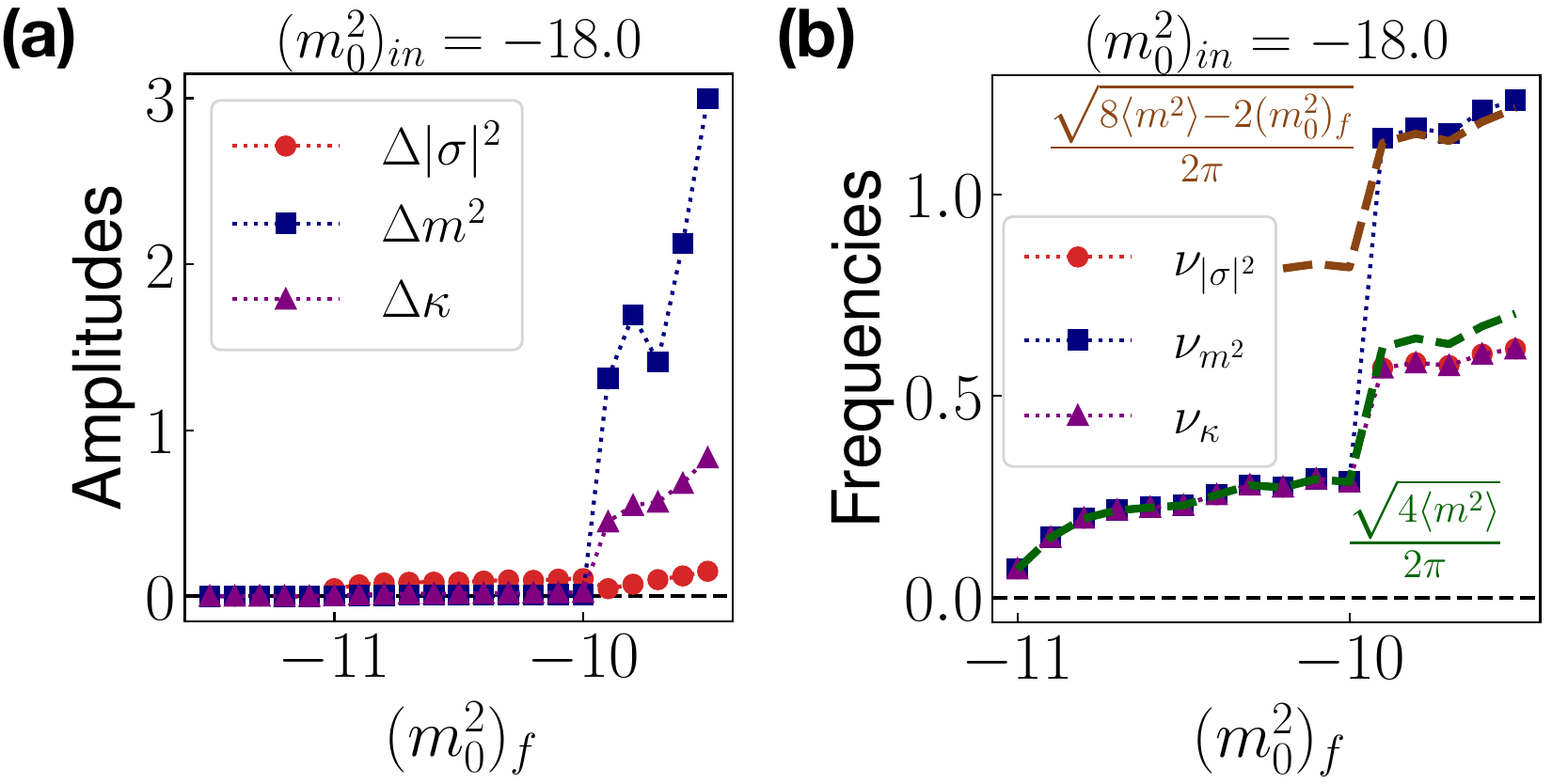}
	\caption[Oscillations of the parameters in the steady state following a sudden quench of $m^2_0$ starting from the condensed phase]{Oscillations of the parameters in the steady state following a sudden quench of $m^2_0$ starting from the condensed phase with $(m^2_0)_{in}=-18.0$ and $\lambda=\lambda_4=2$. $(m^2_0)_{f}$ denotes the final value of $m^2_0$ after the quench. 
		(a) Amplitudes of the oscillations in the steady state. The amplitudes of $m^2$ and $\kappa$ are significantly larger for $(m^2)_f\ge-9.9$. The steady states correspond to the normal phase, where time-averaged $m^2$ is large and $\kappa$ and $|\sigma|^2$ touch zero. 
		(b) Dominant frequencies of the oscillations in the steady state. For $(m^2_0)_f\ge-9.9$, where the oscillations are more prominent, the frequencies agree well with the frequencies that one gets from the dynamical equations. $\nu_{\kappa}$ and $\nu_{|\sigma|^2}$ agree with $\sqrt{4\langle m^2\rangle}/2\pi$ sketched by the dashed green line and  $\nu_{m^2}$ agrees with $\sqrt{8\langle m^2\rangle-(m^2_0)_f}/2\pi$ sketched by the dashed brown line. Here $\langle m^2\rangle$ denotes the time-averaged $m^2$ in the steady state.   
	}
	\label{fracdyn_osc_cond}
\end{figure}
\begin{figure}[!t]
	\centering
	\includegraphics[scale=0.31]{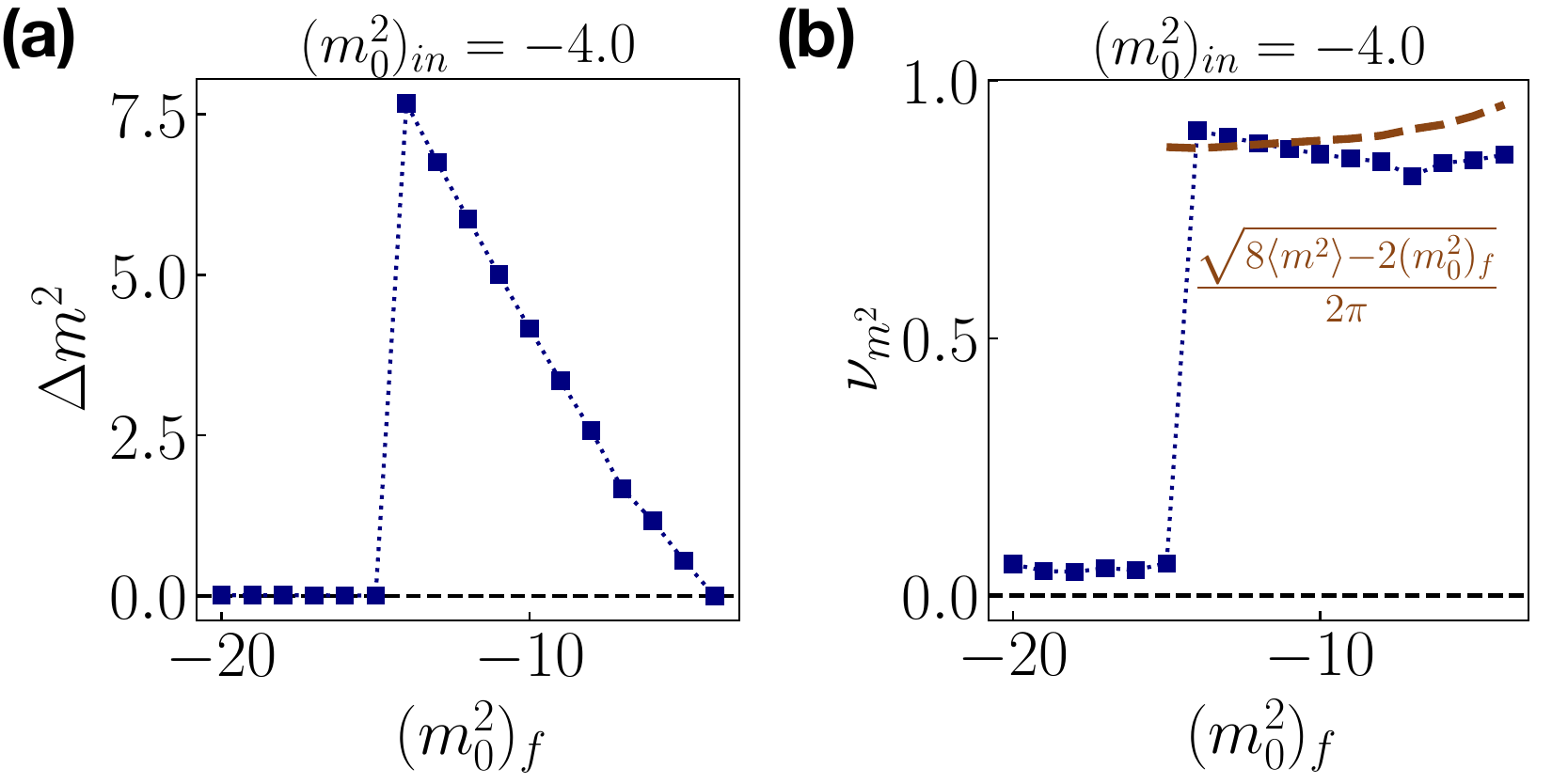}
	\caption[Oscillations of the parameters in the steady state following a sudden quench of $m^2_0$ starting from the normal phase]{Oscillations of the parameter $m^2$ in the steady state following a sudden quench of $m^2_0$ starting from the normal phase with $(m^2_0)_{in}=-4.0$ and $\lambda=\lambda_4=2$. $(m^2_0)_{f}$ denotes the final value of $m^2_0$ after the quench. 
		(a) Amplitudes of the oscillations in the steady state. The amplitudes of $m^2$ and $\kappa$ are significantly larger for $(m^2)_f>-14.9$. The steady states correspond to the normal phase, where time-averaged $m^2$ is finite and $\kappa$ and $|\sigma|^2$ are zero. 
		(b) Dominant frequencies of the oscillations in the steady state. For $(m^2_0)_f>-14.9$, where the oscillations are more prominent, the frequencies agree well with the frequencies that one gets from the dynamical equations. $\nu_{m^2}$ agrees with $\sqrt{8\langle m^2\rangle-(m^2_0)_f}/2\pi$ sketched by the dashed brown line. Here $\langle m^2\rangle$ denotes the time-averaged $m^2$ in the steady state.   
	}
	\label{fracdyn_osc_norm}
\end{figure}
Till now, we focused on the time-averaged values of the parameters in the steady state. However, form Figs.~\ref{quenchcond}(d) and \ref{quenchnorm}(c) it is clear that the parameters have large oscillations for some values of $(m^2_0)_f$, where the time-averaged $\kappa$ is very small and time-averaged $m^2$ is finite and large. In this section, we characterize these oscillations and explain their origin. 

Let us first consider the case of quench from the condensed phase. In Fig~\ref{fracdyn_osc_cond}, the amplitudes and frequencies of the oscillations of the parameters at long times ($t=800-1000$ in the units of $\sqrt[3]{2a^3/\lambda_4}$) are plotted as a function of $(m^2_0)_f$. In Fig.~\ref{fracdyn_osc_cond}(a), the amplitudes of oscillations are very small for $(m^2_0)_f<-9.9$, and they increase significantly for $(m^2_0)_f>-9.9$. This coincides with the jump in $m^2$ and $\kappa$ as shown in Fig.~\ref{quenchcond}(d). 
In Fig.~\ref{fracdyn_osc_cond}(b), the frequencies of  oscillations are plotted. In the region, where the parameters oscillate with large amplitudes, $\kappa$ and $|\sigma|^2$ have the same frequency while $m^2$ oscillates with a different frequency. Both the amplitude and frequency show sharp jumps as the system goes from the intermediate dynamical phase to the normal phase. Similar kind of behaviour is also observed when the system is quenched from the normal phase as shown in Fig.~\ref{fracdyn_osc_norm}. 
In this case only 
$m^2$ shows large oscillations for $(m^2_0)_f>-14.9$, which corresponds to the region in Fig.~\ref{quenchnorm}(c) where steady state value of $m^2$ is finite while $\kappa$ and $|\sigma|^2$ are 0. 
 The frequency of the oscillation also jumps at the same point as can be seen from Fig.~\ref{fracdyn_osc_norm}(b). So, one can conclude that when the time-averaged value of $\kappa$ is close to zero and  time-averaged value of $m^2$ is finite, the parameters show larger and faster oscillations.

To explain these observations, one needs to look back at the dynamical equations depicted in     
Eqs. \ref{meq1}, \ref{dynparams} and \ref{sc1}.
In Eq.~\ref{meq1}, one can make a simple approximation to calculate $G^R(\textbf{k};t,t')$  in the steady state by considering the time averaged values of the parameters $m^2$ and $\kappa$, denoted by $\langle m^2\rangle$ and $\langle \kappa\rangle$ respectively. This results in 
\beq\label{steady_gr}
G^R(\textbf{k};t,t')=-\Theta(t-t')\frac{\sin\langle\omega(\textbf{k})\rangle(t-t')}{2\langle\omega(\textbf{k})\rangle},
\eeq
where $\langle\omega(\textbf{k})\rangle=\sqrt{\langle m^2\rangle+\langle \kappa\rangle\epsilon(\textbf{k})}$. 
Using this and Eq.~\ref{sc1} one gets
\beq\label{steady_gk}
\begin{split}
&G^K(\textbf{k};t,t)\\
&=G^K(\textbf{k};0;0)\left\{\frac{\omega_{in}(\textbf{k})^2}{\langle\omega(\textbf{k})\rangle^2}\sin^2\langle\omega(\textbf{k})\rangle t+\cos^2\langle\omega(\textbf{k})\rangle t\right\}.
\end{split}
\eeq
Re-arranging the time-dependent parts one gets, 
\beq\label{steady_gkt}
\begin{split}
&G^K(\textbf{k};t,t)\\
&=\frac{\langle\omega(\textbf{k})\rangle^2+\omega_{in}(\textbf{k})^2}{4\langle\omega(\textbf{k})\rangle^2\omega_{in}(\textbf{k})}+\frac{\langle\omega(\textbf{k})\rangle^2-\omega_{in}(\textbf{k})^2}{4\langle\omega(\textbf{k})\rangle^2\omega_{in}(\textbf{k})}\cos 2\langle\omega(\textbf{k})\rangle t.
\end{split}
\eeq
Now, from Eq.~\ref{dynparams}, it can be noted that both $m^2(t)$ and $\kappa(t)$ involve integration of $G^K(\textbf{k};t,t)$ over momentum. The time-dependent parts of them involve the integrals
\begin{widetext}
\bseq\label{dyninteg}
\begin{align}
\tilde{I}_1(t)&=\int d\epsilon g(\epsilon)\frac{(\langle m^2\rangle+\langle\kappa\rangle\epsilon)-(m^2_{in}+\kappa_{in}\epsilon)}{4(\langle m^2\rangle+\langle\kappa\rangle\epsilon)\sqrt{m^2_{in}+\kappa_{in}\epsilon}}\cos\left( 2\sqrt{\langle m^2\rangle+\langle\kappa\rangle\epsilon}~ t\right)~~\text{and}\\
\tilde{I}_2(t)&=\int d\epsilon g(\epsilon)\frac{(\langle m^2\rangle+\langle\kappa\rangle\epsilon)-(m^2_{in}+\kappa_{in}\epsilon)}{4(\langle m^2\rangle+\langle\kappa\rangle\epsilon)\sqrt{m^2_{in}+\kappa_{in}\epsilon}}\epsilon\cos\left( 2\sqrt{\langle m^2\rangle+\langle\kappa\rangle\epsilon}~ t\right).
\end{align}
\eseq
Here, we have transformed the momentum integrations to energy integrations using the transformation $\epsilon=\epsilon(\textbf{k})$, where $g(\epsilon)$ is the density of states of the lattice dispersion in 3D.
These are the main time-dependent terms in  $m^2(t)$ and $\kappa(t)$, and their behaviour determines the nature of oscillations in the parameters. 
Note that $\langle\kappa\rangle$ determines the bandwidth of the phase factor. So, large $\langle\kappa\rangle$ causes large dephasing i.e. summing over different phase rotations cancels each other. This results in small $\tilde{I}_1$ and $\tilde{I}_2$ and hence the parameters do not show large oscillations as long as $\langle \kappa\rangle$ stays sufficiently large. On the other hand, a small $\langle \kappa\rangle$ fails to provide enough bandwidth for the phase factors to cancel them. Another way to see this is that performing the integration brings down factors of $\langle \kappa\rangle$ in the denominator and the integrals becomes large when $\langle \kappa\rangle$ is small. As an example let's consider $\tilde{I}_1$ for the case of quench from condensed phase. Considering only the leading order terms one gets
\beq
\tilde{I}_1(t)\sim\int d\epsilon g(\epsilon)\frac{1}{4\sqrt{\kappa_{in}\epsilon}}\cos\left(\sqrt{4\langle m^2\rangle}+\frac{\langle\kappa\rangle}{\sqrt{\langle m^2\rangle}}\epsilon\right)t\sim \frac{\sqrt{\langle m^2\rangle}}{\langle\kappa\rangle}\text{(oscillating term)}.
\eeq
\end{widetext}
So, when  $\langle\kappa\rangle$ is small the integrals $\tilde{I}_1$ and $\tilde{I}_2$ oscillate with large amplitudes and hence one can expect the parameters to have large oscillations. This explains the observed large oscillations of the parameters in the dynamical phase diagrams (see Figs.~\ref{quenchcond}(d) and \ref{quenchnorm}(c)) as shown in Fig.~\ref{fracdyn_osc_cond}(a) and \ref{fracdyn_osc_norm}(a) where  $\langle\kappa\rangle$ is small.

To understand the oscillation frequencies it is useful to take the dynamical equations for the parameters in Eq.~\ref{dynparams} and get differential equations for the parameters out of them. Considering only the simplest terms, one gets
\bseq
\begin{align}
&\partial_t^2\kappa(t)\sim -2m^2(t)\kappa(t)~~\text{and}\\
&\partial_t^2m^2(t) \sim -2[m^2(t)]^2+2(m^2_0)_fm^2(t),
\end{align}
\eseq
respectively.
So, the `instantaneous' oscillation frequencies depends on $m^2(t)$. One can consider time-averaged value of it with a factor of $\sqrt{2}$ to account for the fluctuations and get the oscillation frequencies for $\kappa(t)$ and $m^2(t)$ 
\bseq
\begin{align}
&\tilde{\nu}_\kappa=\sqrt{4\langle m^2\rangle}/2\pi\\
&\tilde{\nu}_{m^2}=\sqrt{8\langle m^2\rangle-2(m^2_0)_f}/2\pi.
\end{align}
\eseq
From Figs.~\ref{fracdyn_osc_cond}(b) and \ref{fracdyn_osc_norm}(b) it can be noted that these frequencies agree well with the observed frequencies in the dynamics of the parameters, when they have large oscillations.

\section{Conclusion}
\label{conc}

We have studied the equilibrium and non-equilibrium phase diagram of
charged interacting bosons with $U(N)$ and dipole symmetries in the
large-$N$ limit. The mass $m^2$, the dispersion $\kappa$ and the
condensate density $|\sigma|^2$ can be used to track the phases of
this system. In thermal equilibrium, the system shows a first-order
phase transition from a phase where both $U(N)$ and dipole symmetry
are intact to a phase where both are broken. Around this transition,
the system has several metastable states of different symmetries and
their energetics leads to a first-order transition.

 In contrast, the steady state of the system starting
from an initial condensed phase and after a quantum quench of the
bare mass parameter $m_0^2$ from $(m_0)^2_{in}$ to $(m_0)^2_f$ shows
a more interesting phase diagram. Here, for a range of $(m_0)^2_f$,
the condensate density goes to zero in a continuous manner. This
indicates a dynamical second order transition. The transition
restores the $U(N)$ symmetry while breaking the dipole symmetry;
this leads to a dipole symmetry broken normal phase which has no
equilibrium counterpart. Upon further increase of $(m_0)^2_f$, the
dipole symmetry broken normal phase gives way to a normal dipole
symmetry restored phase through a first-order phase transition. A
quench from the normal state do not see the intermediate phase; here
we find a direct first-order transition from the dipole symmetry
broken normal to the condensed phase similar to the equilibrium
phase diagram.

Our study of the quantum ramp dynamics elucidates the role of ramp
dynamics in realizing the intermediate phase. We find that the
intermediate phase shrink with increasing $\tau$ and vanishes when
it is above a critical $\tau_c$; this leads to an interpolation
between the quench behaviour $\tau \to 0$ and the adiabatic behaviour
$\tau \to \infty$. 

To conclude, our study shows bosonic systems with dipole symmetry
show interesting equilibrium and non-equilibrium phase diagrams. The
presence of metastable states leads to non-trivial phenomenology
under quench dynamics, where symmetry broken states, which do not
exist in equilibrium, can appear as steady states for certain
initial conditions. These steady states are likely to exists over a
prethermal timescale; this timescale, however, is controlled by
higher order $1/N$ terms in the action and is expected to be large
for large-$N$. They will lead to a slow thermalization in these
systems, a detailed study of which is left as a subject of future
study.

\begin{acknowledgments}
The authors are grateful to Pranay Gorantla and Leticia F. Cugliandolo for useful discussions
and suggestions. MMI and RS acknowledge the use of computational
facilities at the Department of Theoretical Physics, Tata Institute
of Fundamental Research, Mumbai for this paper. MMI and RS
acknowledge support of the Department of Atomic Energy, Government
of India, under Project Identification No. RTI 4002. KS thanks DST,
India for support through SERB project JCB/2021/000030.
\end{acknowledgments}
\begin{widetext}
\appendix
\section{Large-$N$ saddle point equations\label{ApxlargeNsaddle}}
In this appendix we derive the saddle point equations starting from a large-$N$ action for the dipole conserving bosons. We will follow the method described in Ref.~\onlinecite{largeN_fracton}. We start from the euclidean  action (see Eq.~\ref{action_scalar}) in $d$ spatial dimensions
\beq\label{action_scalar_app}
\begin{split}
	S_E=\int d^dx
	\int_0^\beta d\tau~&\left[\phi^{\ast}(x,\tau)(-\partial^2_{\tau}+m_0^2)\phi(x,\tau)+\lambda_4|\phi(x,\tau)|^4\right.\\ 
	&\left.+\lambda\left\{
	~\left(\phi^{\ast}(x,\tau)\right)^2 \left( \phi(x,\tau) \nabla^2
	\phi(x,\tau) -\vec{\nabla} \phi(x,\tau) \cdot \vec{\nabla} \phi(x,\tau)\right)+\text{h.c.}\right\} \right].
\end{split}
\eeq We move to a to a field theory
with $N$ flavours of charged bosons

    \bqa \label{action_scalar_N_apx}
 S_E&=&\int d^dx
 \int_0^\beta d\tau~\left[\sum_a
 \phi^{\ast}_a~(x,\tau)(-\partial^2_{\tau}+m_0^2)\phi_a(x,\tau)+\frac{\lambda_4}{N}\left(\sum_a
 ~ |\phi_a(x,\tau)|^2\right)^2 \right.\\
 \nonumber & &\left. +\frac{\lambda}{N}
 ~\sum_{ab}~\left\{\phi^{\ast}_a(x,\tau)\phi^\ast_b(x,\tau) \left(\frac{1}{2} \left[\phi_a(x,\tau) \nabla^2
 \phi_b(x,\tau)+\phi_b(x,\tau) \nabla^2
 \phi_a(x,\tau)\right] -\vec{\nabla} \phi_a(x,\tau) \cdot \vec{\nabla} \phi_b(x,\tau)\right)+\text{h.c.}\right\} \right],
    \eqa
where $a,b$ denote the flavours of the bosons.

It is convenient to work in the momentum-Matsubara frequency space. We make the Fourier transform $\phi_a(x,\tau)=\int Dk e^{i\textbf{k}.\textbf{x}-i\omega_n\tau}\phi_a(k)$.
The action in momentum-Matsubara frequency space is
\beq\label{fracton_scalar_app}
    S_E=\int Dk~\sum_{a}~\phi^{*}_a(k)\left(\omega^2_n+m^2_0\right)\phi_a(k)+\int \prod_{i=1}^4Dk_i~\sum_{ab}~\phi^{*}_a(k_1)\phi^{*}_b(k_2)\phi_b(k_3)\phi_a(k_4)V(\{k_i\}),
\eeq
where $k=(\textbf{k},\omega_n)$, $\int Dk=\frac{1}{\beta}\sum_{n}\int\frac{d^dk}{(2\pi)^d}$ and $V(\{k_i\})=\left[\frac{\lambda_4}{N}+\frac{\lambda}{2N}\left\{|\textbf{k}_1-\textbf{k}_2|^2+|\textbf{k}_3-\textbf{k}_4|^2\right\}\right]\delta(k_1+k_2-k_3-k_4)$.

In the path integral we introduce an resolution of identity in terms of the correlator $G(k,k')=\sum_{a}\langle\phi^{*}_a(k)\phi_{a}(k')\rangle$ and a Lagrange multiplier $\Sigma(k,k')$ as
$\mathbb{I} =\int DG D\Sigma~ e^{\int DkDk'~\Sigma(k,k')\left[G(k,k')-\sum_{a}\langle\phi^{*}_a(k)\phi_{a}(k')\rangle\right]}$.
With this the large-$N$ action takes the form
\beq\label{largeN_action}
\begin{split}
S_N&=\int Dk~\sum_a~\phi^{*}_a(k)\left[\omega^2_n+\Sigma(k,k)\right]\phi_a(k)+m^2_0\int Dk~G(k,k)\\&-\int Dk_1Dk_2\Sigma(k_1,k_2)G(k_1,k_2)+\int \prod_{i=1}^4Dk_i~G(k_1,k_3)G(k_2,k_4)V(\{k_i\})
\end{split}
\eeq
Integrating out $(N-1)$ fields and setting $\sigma(k)=\langle\phi_{a=1}(k)\rangle/\sqrt{N}$, we get
\beq
\begin{split}
    S'_N&=(N-1)\int Dk \ln\left[\omega^2_n+\Sigma(k,k)\right]+N\int Dk~\sigma^{*}(k)\left(\omega^2_n+\Sigma(k,k)\right)\sigma(k)+m^2_0\int Dk~G(k,k)\\&-\int Dk_1Dk_2\Sigma(k_1,k_2)G(k_1,k_2)+\int \prod_{i=1}^4Dk_i~G(k_1,k_3)G(k_2,k_4)V(\{k_i\}).
\end{split}
\eeq
By assuming translational invariance we have,
\beq
\begin{split}
    G(k,k')&=N\beta\delta_{n_1n_2}(2\pi)^d\delta^d(\textbf{k}-\textbf{k}')G(k')\\
    \Sigma(k,k')&=\beta\delta_{n_1n_2}(2\pi)^d\delta^d(\textbf{k}-\textbf{k}')\Sigma(k')~\text{and}\\
    \sigma(k)&=\beta\delta_{n0}(2\pi)^d\delta^d(\textbf{k})\sigma,
\end{split}
\eeq
Putting all of these we get the effective action
\beq\label{action_eff}
\begin{split}
        S'_N&=(N-1)\int Dk \ln\left[\omega^2_n+\Sigma(k)\right]+N\beta\delta_{n0}(2\pi)^d\delta^d(\textbf{k})|\sigma|^{2}\Sigma(k=0)+Nm^2_0\int Dk~G(k)\\&-N\int Dk\Sigma(k)G(k)+N\int Dk Dk'~NV(\textbf{k},\textbf{k}';\textbf{k},\textbf{k}')G(k)G(k').
\end{split}
\eeq
Taking variation of the action with respect to $\Sigma$, $G$ and $\sigma$ we get the saddle point equations
\beq
\begin{split}
    G(k)&=\frac{1}{\omega_n^2+\Sigma(k)}+\beta\delta_{n0}(2\pi)^d\delta^d(\textbf{k})|\sigma|^{2}\\
    \Sigma(k)&=m^2_0+2\int Dk'~NV(\textbf{k},\textbf{k}';\textbf{k},\textbf{k}')G(k')~\text{and}\\
    &\sigma\Sigma(\textbf{k}=0)=0.
\end{split}
\eeq
It is clear that $\Sigma(k)$ only depends on the spatial component $\textbf{k}$ of the space-time vector $k=(\textbf{k},\omega_n)$ and it will represent square of the effective dispersion of the mode $\textbf{k}$. So we set $\Sigma(k)=\omega(\textbf{k})^2$. Final set of equations are given in Eq.~\ref{largeN_dyson}.
From Eq.~\ref{largeN_dyson_a}  we can put the expression for $G(k)$ in Eq.~\ref{largeN_dyson_b} to get
\beq\label{selcon_derive}
\omega(\textbf{k})^2=m^2_0+2|\sigma|^2NV(\textbf{k},0;\textbf{k},0)+2\int\frac{d^dk'}{(2\pi)^d}NV(\textbf{k},\textbf{k}';\textbf{k},\textbf{k}')\frac{1}{\beta}\sum_{n}\frac{1}{\omega_n^2+\omega(\textbf{k}')^2}
\eeq
Now the Matsubara sum becomes
\beq
\begin{split}
    \frac{1}{\beta}\sum_{n}\frac{1}{\omega_n^2+\omega(\textbf{k}')^2}&=-\frac{1}{\beta}\sum_{n}\frac{1}{(i\omega_n)^2-\omega(\textbf{k}')^2}\\
    &=-\frac{1}{2\omega(\textbf{k}')}\frac{1}{\beta}\sum_{n}\left[\frac{1}{i\omega_n-\omega(\textbf{k}')}-\frac{1}{i\omega_n+\omega(\textbf{k}')}\right]\\
    &=\frac{1}{2\omega(\textbf{k}')}\left[n_B\left(\omega(\textbf{k}')\right)-n_B\left(-\omega(\textbf{k}')\right)\right],
\end{split}
\eeq
where $n_B(\omega)=\frac{1}{e^{\beta\omega}-1}$. So we get $    \frac{1}{\beta}\sum_{n}\frac{1}{\omega_n^2+\omega(\textbf{k}')^2}=\frac{1}{2\omega(\textbf{k}')}\left[1+2n_B\left(\omega(\textbf{k}')\right)\right]=\frac{\coth\left(\frac{\omega(\textbf{k}')}{2T}\right)}{2\omega(\textbf{k}')}$. Putting this in Eq.~\ref{selcon_derive}, we get self-consistent equation for the dressed dispersion $\omega(\textbf{k})$ (presented in Eq.~\ref{largeN_selfcon} in the main text)
\beq
\omega(\textbf{k})^2=m^2_0+2|\sigma|^2NV(\textbf{k},0;\textbf{k},0)+2\int\frac{d^dk'}{(2\pi)^d}NV(\textbf{k},\textbf{k}';\textbf{k},\textbf{k}')\frac{\coth\left(\frac{\omega(\textbf{k}')}{2T}\right)}{2\omega(\textbf{k}')}.
\eeq

To regulate ultraviolet divergences we work on a lattice and use the regularized interaction (see Eq.~\ref{vertex_lattice}) $NV_r(\textbf{k},\textbf{k}';\textbf{k},\textbf{k}')=\lambda\epsilon(\textbf{k}-\textbf{k}')+\lambda_4$, where $\epsilon(\textbf{k}-\textbf{k}')=\frac{4}{a^2}\sum_{i=1}^{d}\sin^2\left(\frac{(k_i-k'_i)a}{2}\right)$ for a hyper cubic lattice in $d$ dimension with lattice constant $a$. Using the structure of the interaction vertex, we can make the following ansatz for the dressed dispersion
\beq
\omega(\textbf{k})^2=m^2+\kappa\epsilon(\textbf{k}),
\eeq
where $m^2$ is the gap and $\kappa$ is related to the curvature or effective mass in kinetic sense. Putting this in Eq.~\ref{largeN_selfcon} we get
\beq\label{selfcon_der1}
m^2+\kappa\epsilon(\textbf{k})=m^2_0+2|\sigma|^2\left[\lambda_4+\lambda\epsilon(\textbf{k})\right]+2\int\frac{d^dk'}{(2\pi)^d}\left[\lambda_4+\lambda\epsilon(\textbf{k}-\textbf{k}')\right]\frac{\coth\left(\frac{\omega(\textbf{k}')}{2T}\right)}{2\omega(\textbf{k}')}.
\eeq
Simplifying we get
\beq\label{selfcon_der2}
m^2+\kappa\epsilon(\textbf{k})=m^2_0+2|\sigma|^2\left[\lambda_4+\lambda\epsilon(\textbf{k})\right]+2\int\frac{d^dk'}{(2\pi)^d}\left[\left\{\lambda_4+\lambda\epsilon(\textbf{k}')\right\}+\lambda\left\{1-\frac{a^2}{2d}\epsilon(\textbf{k}')\right\}\epsilon(\textbf{k})\right]\frac{\coth\left(\frac{\omega(\textbf{k}')}{2T}\right)}{2\omega(\textbf{k}')}.
\eeq
Equating co-efficient of $\epsilon(\textbf{k})$ and $\textbf{k}$-independent terms on both sides of Eq.~\ref{selfcon_der2} we get
\begin{subequations}
    \begin{align}
        m^2&=m^2_0+2|\sigma|^2\lambda_4+2\lambda_4I_1(m^2,\kappa)+2\lambda I_2(m^2,\kappa)~\text{and}\\
        \kappa&=2|\sigma|^2\lambda+2\lambda [I_1(m^2,\kappa)-(a^2/2d) I_2(m^2,\kappa)],
    \end{align}
\end{subequations}
where $I_1(m^2,\kappa)=\int\frac{d^dk}{(2\pi)^d}\frac{\coth\left(\frac{\omega(\textbf{k})}{2T}\right)}{2\omega(\textbf{k})}$ and $I_2(m^2,\kappa)=\int\frac{d^dk}{(2\pi)^d}\frac{\epsilon(\textbf{k})\coth\left(\frac{\omega(\textbf{k})}{2T}\right)}{2\omega(\textbf{k})}$. We also have to satisfy $\sigma\Sigma(k)=\sigma\omega^2(\textbf{k})=0$, which with our ansatz translates to $\sigma m^2=0$. This means either of $\sigma$ and $m^2$ has to be zero. Setting $\sigma=0$ we get the equations for the normal phase depicted in Eq.~\ref{normeqs} for $d=3$. These equations represent a phase which conserve charge. The other set of equations can be found by setting $m^2=0$. These equations represent a condensed phase which breaks charge and consequently dipole symmetry and are presented in Eq.~\ref{condeqs}.

\section{Phases and transition with varying coupling at T=0\label{ApxPhaseSol}}
\begin{figure*}[h]
    \includegraphics[scale=0.36]{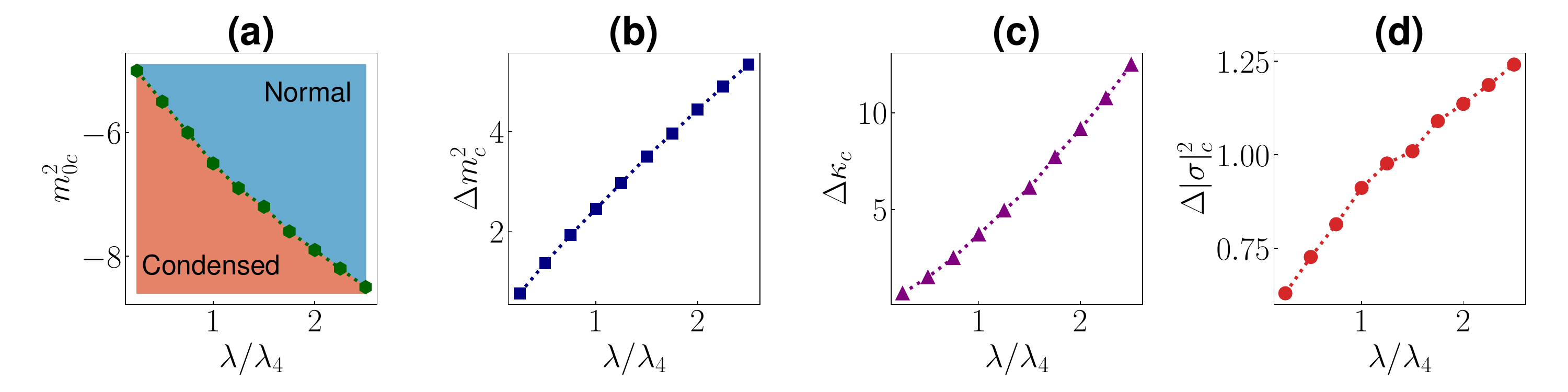}
    \caption{Phases and transition at $T=0$ for varying couplings. We keep $\lambda_4=2$ fixed and vary $\lambda$. (a) Critical $m^2_{0}$ is plotted as a function of $(\lambda/\lambda_4)$. It decreases with $(\lambda/\lambda_4)$. The curve also represent the boundary between condensed phase (shaded in light red) and normal phase (shaded in light blue). (b)-(d) Jumps in the phase parameters at the transition are plotted with varying couplings. The jumps increases with $(\lambda/\lambda_4)$ keeping the transition first-order in nature. }
    \label{phases_T0difflamb}
\end{figure*}
In this appendix we present more results about the phases and transition at equilibrium. In the main text we kept the couplings fixed at $\lambda_4=2$ and $\lambda=2$ and we observed a discontinuous transition between a condensed phase ($U(N)$ and dipole symmetry broken) and a normal phase ($U(N)$ and dipole symmetry conserved). Here we investigate the phase diagram by changing $\lambda$ while keeping $\lambda_4=2$ fixed. We find qualitatively similar behaviour here. In Fig.~\ref{phases_T0difflamb}(a), we present the transition mass squared $m^2_{0c}$ as a function $(\lambda/\lambda_4)$. We see that $m^2_{0c}$ decreases as $(\lambda/\lambda_4)$ is increased. This plot also represent the phase diagram in $(\lambda/\lambda_4)-m^2_0$ plane. The curve for $m^2_{0c}$ separates the two phases: condensed (shaded in light red) and normal (shaded in light blue). In Fig.~\ref{phases_T0difflamb}(b)-(d), we plot the absolute value of the jumps in the phase parameters $m^2$, $\kappa$ and $|\sigma|^2$ respectively at the transition. The jumps increases with increasing $(\lambda/\lambda_4)$ for all the parameters keeping the transition first-order in nature.
\section{Phase parameters at thermal equilibrium\label{ApxPhaseTh}}
\begin{figure*}[h!]
    \includegraphics[scale=0.38]{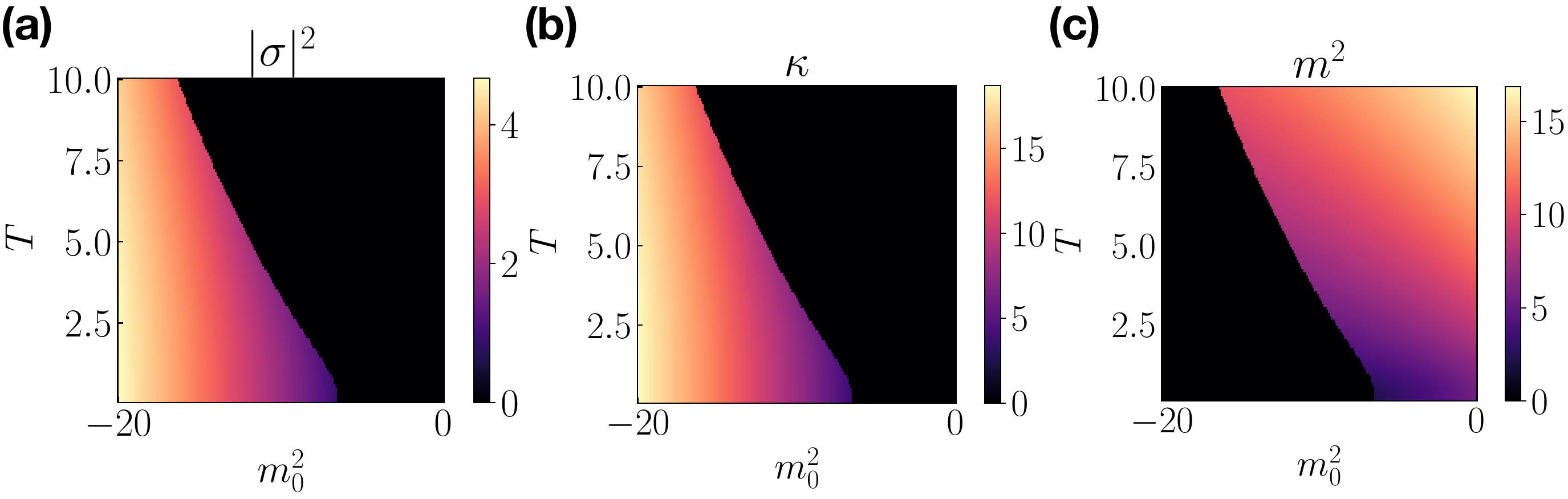}
    \caption{Phase parameters of dipole conserving bosons at thermal equilibrium. (a), (b) and (c) Colormap of $|\sigma|^2$, $\kappa$ and $m^2$ respectively on the $m^2_0-T$ plane. The black regions represent 0 values. The coloured regions represent finite values given by the accompanying colorbars. The boundary between these two regions represent a first- order transition. $|\sigma|^2$ and $\kappa$ are finite in the same region which hosts a condensed phase. On the other hand $m^2$ is finite in a region where $|\sigma|^2=0$ and $\kappa=0$. This represents a normal phase. We also note that the phase parameters changes discontinuously at the transition which is the signature of the fist order nature of the transition.      }
    \label{phases_thermal_apx}
\end{figure*}
In this appendix we present the phase parameters at thermal equilibrium. Specifically we plot them as colormap on $m^2_0-T$ to see their variation with the bare mass and temperature. Their values (finite or not) determines the phase.
In Fig.~\ref{phases_thermal_apx} we present the colormaps of the phase parameters in $m^2_0-T$ plane corresponding to the phase diagram presented in Fig.~\ref{quenchcond}(e). In this figures, the black regions represent 0 values. The coloured regions represent finite values given by the accompanying colorbars. The boundary between these two regions represent a first-order transition. $|\sigma|^2$ and $\kappa$ are finite in the same region which hosts a condensed phase. On the other hand $m^2$ is finite in a region where $|\sigma|^2=0$ and $\kappa=0$. This represents a normal phase. We note that the phase parameters changes discontinuously at the transition which is the signature of the fist order nature of the transition. We also note that the critical $m^2_0$ decreases with temperature.

\section{Dynamical equations for parameters\label{ApxDynEq}}
In this appendix, we present the details of the method for the non-equilibrium dynamics of dipole conserving bosons. First, we derive the dynamical equations for the parameters in terms of one-body correlators. The correlators are evolved following SKFT for scalar field from Fock states\cite{realscalar} and described towards the end of this appendix.
We initialize the system in the ground state of a particular phase with a fixed values of the phase parameters ($m^2_{in}$,$\kappa_{in}$ and $|\sigma|^2_{in}$) corresponding to $m^2_0=(m^2_{0})_{in}$. The initial Keldysh correlator is related to these parameters as
\beq
iG^K(\textbf{k};t=0,t'=0)=
\begin{cases}
\frac{1}{2\sqrt{m^2_{in}+\kappa_{in}\epsilon(\textbf{k})}} & \text{for}~\textbf{k}\ne 0\\
(2\pi)^d|\sigma|^2_{in}& \text{for}~\textbf{k}=0.
\end{cases}
\eeq

Now we proceed to describe how to calculate the phase parameters for the subsequent time
in the case of time-dependent bare mass $m^2_0(t)$.
 Within the SKFT structure\cite{kamenevbook}, the effective dispersion at any time is related to the retarded self-energy as
\beq\label{dispsr}
\omega^2(\textbf{k},t)=m^2_0(t)+\frac{1}{2}\Sigma^R(\textbf{k},t).
\eeq
For the dynamics, the large-$N$ (mean field) self-energy is given by the Hartree diagram \cite{kamenevbook}. 
The retarded self-energy is given in terms of the interaction vertex and Keldysh correlator. To get the interaction vertex in Keldysh basis we write the interacting part of the action in terms of the fields in Keldysh basis as 
\beq
\begin{split}
&S_{int}=-\int \frac{d^dk}{(2\pi)^d}\int \frac{d^dk'}{(2\pi)^d}\int_{0}^{\infty} dt~ V_r(\textbf{k},\textbf{k}';\textbf{k},\textbf{k}')\left[\phi^{*}_{+}(\textbf{k},t)\phi^{*}_{+}(\textbf{k}',t)\phi_{+}(\textbf{k},t)\phi_{+}(\textbf{k}',t)-\phi^{*}_{-}(\textbf{k},t)\phi^{*}_{-}(\textbf{k}',t)\phi_{-}(\textbf{k},t)\phi_{-}(\textbf{k}',t)\right]\\
&=-\int \frac{d^dk}{(2\pi)^d}\int \frac{d^dk'}{(2\pi)^d}\int_{0}^{\infty} dt~ V_{K}(\textbf{k},\textbf{k}')\left[\phi^{*}_{cl}(\textbf{k},t)\phi^{*}_{cl}(\textbf{k}',t)\phi_{cl}(\textbf{k},t)\phi_{q}(\textbf{k}',t)+\phi^{*}_{cl}(\textbf{k},t)\phi^{*}_{q}(\textbf{k}',t)\phi_{cl}(\textbf{k},t)\phi_{cl}(\textbf{k}',t)\right.\\
&~~~~~~~~~~~~~~~~~~~~~~~~~~~~~~~~~~~~~~~~~~~~~~~~~~~~~~~~~~\left.+\phi^{*}_{q}(\textbf{k},t)\phi^{*}_{q}(\textbf{k}',t)\phi_{q}(\textbf{k},t)\phi_{cl}(\textbf{k}',t)+\phi^{*}_{q}(\textbf{k},t)\phi^{*}_{cl}(\textbf{k}',t)\phi_{q}(\textbf{k},t)\phi_{q}(\textbf{k}',t)\right],
\end{split}
\eeq
where the vertex  $V_{K}(\textbf{k},\textbf{k}')=4\lambda\epsilon(\textbf{k}-\textbf{k}')+4\lambda_4=4V_r(\textbf{k},\textbf{k}';\textbf{k},\textbf{k}')$. Going to the $2^{nd}$ line from the $1^{st}$ line, we have used the Keldysh rotated fields namely classical fields $\phi_{cl}=\frac{1}{2}(\phi_{+}+\phi_{-})$ and quantum fields $\phi_{q}=\frac{1}{2}(\phi_{+}-\phi_{-})$ \cite{kamenevbook}.
Using this vertex we get the retarded self-energy
\beq\label{sr_keldysh}
\Sigma^R(\textbf{k},t)=\int \frac{d^dk'}{(2\pi)^d}V_{K}(\textbf{k},\textbf{k}')iG^K(\textbf{k}';t,t)+V_{K}(\textbf{k},0)|\sigma|^2(t),
\eeq
where $|\sigma|^2(t)=iG^K(\textbf{k}=0;t,t)/(2\pi)^d$.
For the effective dispersion (Eq.~\ref{dispsr}) we get
\beq\label{dispdyn}
\omega^2(\textbf{k},t)=m^2_0(t)+2\left[\lambda_4+\lambda\epsilon(\textbf{k})\right]|\sigma|^2(t)+2\int\frac{d^dk'}{(2\pi)^d}\left[\left\{\lambda_4+\lambda\epsilon(\textbf{k}')\right\}+\lambda\left\{1-\frac{a^2}{2d}\epsilon(\textbf{k}')\right\}\epsilon(\textbf{k})\right]iG^K(\textbf{k}';t,t)
\eeq
Using the parametrization $\omega^2(\textbf{k},t)=m^2(t)+\kappa(t)\epsilon(\textbf{k})$ in analogy to the equilibrium case and comparing co-efficient of $\epsilon(\textbf{k})$ and $\textbf{k}$-independent terms in Eq.~\ref{dispdyn} we get
\bseq
\begin{align}
    m^2(t)&=m_0^2(t)+2\lambda_4|\sigma|^2(t)+2\lambda_4\int\frac{d^dk}{(2\pi)^d}iG^K(\textbf{k};t,t)+2\lambda\int\frac{d^dk}{(2\pi)^d}\epsilon(\textbf{k})iG^K(\textbf{k};t,t)~\text{and}\\
    \kappa(t)&=2\lambda|\sigma|^2(t)+2\lambda\int\frac{d^dk}{(2\pi)^d}iG^K(\textbf{k};t,t)-2\lambda(a^2/2d)\int\frac{d^dk}{(2\pi)^d}\epsilon(\textbf{k})iG^K(\textbf{k};t,t).
\end{align}
\eseq

Now we describe the time-evolutions of one-body correlators. Dynamics of the correlators are required to get the time-evolution of the parameters.
We are concerned about the dynamics of the one-body correlators of the SKFT structure, namely the retarded correlator $iG^R(\textbf{k};t,t')=\langle\phi_{cl}(\textbf{k},t)\phi^*_{q}(\textbf{k},t')\rangle$ and Keldysh correlator $iG^K(\textbf{k};t,t')=\langle\phi_{cl}(\textbf{k},t)\phi^*_{cl}(\textbf{k},t')\rangle$. Within the mean field approximation, there is only Hartree diagram for retarded self-energy and there is no Keldysh self-energy. If we start from a Fock state with mode energies $\omega_{in}(\textbf{k})$, the correlators are determined by
\bseq
\begin{align}
    &2\left[-\partial_t^2 -m^2(t) -\kappa(t)
    \epsilon(\textbf{k})\right] G^R(\textbf{k},t,t')= \delta(t-t')~
    \text{and}\\
    G^K(\textbf{k};t,t)&=[2\omega_{in}(\textbf{k})]^2G^K(\textbf{k};0,0)\left\{|G^R(\textbf{k};t,0)|^2+\frac{1}{[\omega_{in}(\textbf{k})]^2}|\bar{G}^{R}(\textbf{k};t,0)|^2\right\},
\end{align}
\eseq
where  
$\bar{G}^R(\textbf{k};t,t')=\partial_{t'}G^R(\textbf{k};t,t')$.
So, we now have closed set of equations for the dynamics of the parameters $m^2(t)$, $\kappa(t)$ and $|\sigma|^2(t)$ and the correlators $G^R(\textbf{k};t,t')$ and $G^K(\textbf{k};t,t)$.
\end{widetext}
\bibliography{fracdyn_refs.bib}
\end{document}